\documentclass[aps,prd,preprint,superscriptaddress,showpacs]{revtex4}
\usepackage{graphicx}

\usepackage{ulem}
\usepackage{color}
\definecolor{My_red}        {cmyk}{0.00,1.00,1.00,0.20}

\newcommand{\bmat}{\left(\begin{array}}
\newcommand{\emat}{\end{array}\right)}
\newcommand{\beq}{\begin{equation}}
\newcommand{\eeq}{\end{equation}}

\def\nnb{\nonumber}

\def\bwt{\begin{widetext}}
\def\ewt{\end{widetext}}
\def\be{\begin{equation}}
\def\ee{\end{equation}}
\def\bea{\begin{eqnarray}}
\def\eea{\end{eqnarray}}
\def\bean{\begin{eqnarray*}}
\def\eean{\end{eqnarray*}}
\def\bary{\begin{array}}
\def\eary{\end{array}}
\def\bit{\begin{itemize}}
\def\eit{\end{itemize}}

\def\su5u1{SU(5) \times U(1)}
\def\fsu5u1{SU(5) \times U(1)'}
\def\so10{SO(10)}
\def\sq20{SO(10) \times SO(10)}

\def\bwt{\begin{widetext}}
\def\ewt{\end{widetext}}
\def\be{\begin{equation}}
\def\ee{\end{equation}}
\def\bea{\begin{eqnarray}}
\def\eea{\end{eqnarray}}
\def\bean{\begin{eqnarray*}}
\def\eean{\end{eqnarray*}}
\def\bary{\begin{array}}
\def\eary{\end{array}}
\def\bit{\begin{itemize}}
\def\eit{\end{itemize}}

\def\su5u1{SU(5) \times U(1)}
\def\fsu5u1{SU(5) \times U(1)'}
\def\so10{SO(10)}
\def\sq20{SO(10) \times SO(10)}

\usepackage[centertags]{amsmath}
\usepackage{amssymb}

\begin{document}

\title{Flipped $SU(5)\times U(1)_X$ Models from F-Theory}

\author{Jing Jiang}

\affiliation{Department of Physics, University of Wisconsin, 
Madison, WI 53706, USA}

\author{Tianjun Li}

\affiliation{George P. and Cynthia W. Mitchell Institute for
Fundamental Physics, Texas A$\&$M University, College Station, TX
77843, USA }

\affiliation{Key Laboratory of Frontiers in Theoretical Physics, 
      Institute of Theoretical Physics, Chinese Academy of Sciences, 
Beijing 100190, P. R. China }

\author{Dimitri V. Nanopoulos}

\affiliation{George P. and Cynthia W. Mitchell Institute for
Fundamental Physics,
 Texas A$\&$M University, College Station, TX 77843, USA }

\affiliation{Astroparticle Physics Group,
Houston Advanced Research Center (HARC),
Mitchell Campus, Woodlands, TX 77381, USA}

\affiliation{Academy of Athens, Division of Natural Sciences,
 28 Panepistimiou Avenue, Athens 10679, Greece }

\author{Dan Xie}

\affiliation{George P. and Cynthia W. Mitchell Institute for
Fundamental Physics, Texas A$\&$M University, College Station, TX
77843, USA }

\date{\today}

\begin{abstract}

We systematically construct flipped $SU(5)\times U(1)_X$ 
models without and with bulk vector-like particles from
F-theory. To realize the decoupling scenario,  we introduce 
sets of vector-like particles in complete $SU(5)\times U(1)$ 
multiplets at the TeV scale, or at the intermediate scale, 
or at the TeV scale and high scale. To avoid the Landau 
pole problem for the gauge couplings, we can only introduce 
five sets of vector-like particles around the TeV scale. 
These vector-like particles can couple to the Standard 
Model singlet fields, and obtain suitable masses by Higgs 
mechanism. We study gauge coupling unification in detail. 
We show that the $U(1)_X$ flux contributions to the gauge
couplings preserve the $SU(5)\times U(1)_X$ gauge coupling
unification. We calculate the $SU(3)_C\times SU(2)_L$ unification 
scales, and the $SU(5)\times U(1)_X$ unification scales
and unified couplings.  In most of our models, the 
high-scale or bulk vector-like particles can be 
considered as string-scale threshold corrections 
since their masses are close to the string scale.
Futhermore, we discuss the phenomenological consequences 
of our models. In particular, in the models with TeV-scale 
vector-like particles, the vector-like particles can be 
observed at the Large Hadron collider, the proton decay 
is within the reach of the future Hyper-Kamiokande 
experiment,  the lightest CP-even Higgs boson mass 
can be increased, the hybrid inflation can be naturally 
realized, and the correct cosmic primodial density fluctuations 
can be generated.

\end{abstract}


\preprint{ACT-06-09, MIFP-09-23}

\maketitle

\section{Introduction}

The goal of string phenomenology is to construct realistic
string models with moduli stabilization and without chiral
exotics, and then make clean predictions that can tested at
the Large Hadron Collider (LHC) and other future experiments. 
As we know, there are three kinds of string models which have 
been studied extensively 

(1) Heterotic $E_8\times E_8$ string model building. The
supersymmetric Standard Model (SM) can be constructed via
the orbifold 
compactifications~\cite{Buchmuller:2005jr, Lebedev:2006kn, Kim:2006hw} 
and the Calabi-Yau manifold 
compactifications~\cite{Braun:2005ux, Bouchard:2005ag}.
The orbifold compactifications are based on the weakly coupled
heterotic $E_8\times E_8$ string theory, and the Minimal 
Supersymmetric Standard Model (MSSM) without chiral exotic 
particles can be constructed~\cite{Buchmuller:2005jr, Lebedev:2006kn}. 
However,  the gauge coupling 
unification scale in the MSSM is 
around $2\times 10^{16}$ GeV~\cite{Langacker:1991an}, while
 the string scale $M_{\rm string}$ in the weakly coupled heterotic 
string theory is~\cite{Dienes:1996du}
\begin{eqnarray}
M_{\rm string} = g_{\rm string} \times 5.27 \times 10^{17} ~{\rm GeV}~,~\,
\label{String-U}
\end{eqnarray}
where $g_{\rm string}$ is the string coupling constant. Note that
$g_{\rm string} \sim {\cal O} (1)$, we have
\begin{eqnarray}
M_{\rm string} \approx 5 \times 10^{17} ~{\rm GeV}~.~\,
\end{eqnarray}
Thus, there exists a factor of approximately 20 to 25 between 
 the MSSM unification scale and the string scale. 
This problem can be solved in the strong
coupled heterotic $E_8\times E_8$ string theory or M-theory
on $S^1/Z_2$~\cite{Horava:1995qa} 
with Calabi-Yau manifold compactifications 
since the eleventh dimension can be relatively  
large about $10^{14}~{\rm GeV}$~\cite{Witten:1996mz},
and the Grand Unified Theories (GUTs) can be 
realized~\cite{Braun:2005ux, Bouchard:2005ag}.
To break the GUT group via the Wilson line mechanism, the
fundamental group of the Calabi-Yau manifolds should be 
non-trivial. Although the desirable Calabi-Yau manifolds
can be constructed~\cite{Braun:2005ux, Bouchard:2005ag},  
there do exist the following
problems: the vanishing down-type quark Yukawa couplings;
the possible R-parity violations; and the extra massless
$U(1)$ if the rank of GUT group is five or higher.

(2) Free fermionic string model builing. Realistic models
with clean particle spectra can only be constructed at 
the Kac-Moody level one~\cite{AEHN, Faraggi:1989ka,
Antoniadis:1990hb, LNY, Cleaver:2001ab}. Note that the Higgs
fields in the adjoint representation or higher can not be
generated at the Kac-Moody level one, only three kinds
of models can be constructed: the Standard-like models,
Pati-Salam models, and flipped $SU(5)$ 
models~\cite{AEHN, Faraggi:1989ka,
Antoniadis:1990hb, LNY, Cleaver:2001ab}.

(3) D-brane model building. There are two kinds of such
models: (i) Intersecting D-brane models~\cite{Berkooz:1996km,
Ibanez:2001nd, Blumenhagen:2001te, CSU, Cvetic:2002pj, 
CLL, Chen:2005ab, Chen:2005mj, Blumenhagen:2005mu}; (ii) 
Orientifolds of Gepner 
models~\cite{Dijkstra:2004ym, Dijkstra:2004cc}. 
The standard-like models, 
Pati-Salam models, trinification models,
$SU(5)$ models, as well as flipped $SU(5)$ models
have been constructed~\cite{Ibanez:2001nd, 
Blumenhagen:2001te, CSU, Cvetic:2002pj, CLL, Chen:2005ab, 
Chen:2005mj, Dijkstra:2004ym, Dijkstra:2004cc}. However, 
in the  trinification models,
$SU(5)$ models, and flipped $SU(5)$ models,
some of the SM fermion 
Yukawa couplings are forbidden due to the
remaining global $U(1)$ symmetries at 
the perturbative level, for example, 
 the up-type quark Yukawa couplings 
$\mathbf{10_i 10_j 5_{H}}$ in the $SU(5)$ model.
This problem might be solved in the Type IIB
orientifold compactifications~\cite{Blumenhagen:2008zz} 
due to non-perturbative instanton 
effects~\cite{Blumenhagen:2006xt}. 
In the standard-like models and
Pati-Salam models, we can have all the SM fermion 
Yukawa couplings at 
the stringy tree level in principle. However, there are 
some problems in the generic standard-like models 
and Pati-Salam models: rank-one problem in the SM 
fermion Yukawa coupling matrices,
no gauge coupling unification, and additional exotic 
particles, etc. These problems can be solved only in 
a few models~\cite{Chen:2007px, Chen:2007zu}.

On the other hand, 
there are strong indications favoring GUTs from the known 
low-energy particle physics. The gauge 
couplings in the MSSM are indeed unified at the
GUT scale $M_{\rm GUT}$ around $2\times 10^{16}$ 
GeV~\cite{Langacker:1991an}.
Moreover, one family of the SM fermions forms the 
${\mathbf{10}}$ and ${\mathbf{\overline{5}}}$ 
representations in $SU(5)$ models and a single
spinor ${\mathbf{16}}$ representation in 
$SO(10)$ models. Especially, we indeed 
can have the Yukawa coupling
unification for the third family of 
the SM fermions~\cite{Gogoladze:2005az}. 
Also, GUTs can explain charge quantization naturally, etc. 
Therefore, it is very interesting to 
construct GUTs especially $SO(10)$ models from 
the string theory. 

Recently, semi-realistic GUTs have been constructed locally in 
the F-theory with seven-branes, which can be considered as the
strongly coupled formulation of ten-dimensional Type IIB string 
theory with a varying axion ($a$)-dilaton ($\phi$) field 
$\tau=a+ie^{-\phi}$~\cite{Vafa:1996xn, Donagi:2008ca,
Beasley:2008dc, Beasley:2008kw, Donagi:2008kj}
(For a briefly review, see Section III.). Then further model
building and phenomenological consequences have been studied 
extensively~\cite{Heckman:2008es, Marsano:2008jq, Marsano:2008py,
Heckman:2008qt, Font:2008id, Braun:2008pz, Heckman:2008qa,
Jiang:2008yf, Collinucci:2008zs, Blumenhagen:2008aw, Heckman:2008jy,
Bourjaily:2009vf, Hayashi:2009ge, Andreas:2009uf, Chen:2009me,
Heckman:2009bi, Donagi:2009ra, Bouchard:2009bu, Randall:2009dw,
Heckman:2009de, Marsano:2009ym, Bourjaily:2009ci}. 
Note that the known GUTs without additional chiral 
exotic particles are asymptotically free, and asymptotic 
freedom can be translated into the existence of a consistent 
decompactification limit. Also, the Planck scale $M_{\rm Pl}$ 
is about $10^{19}$ GeV, so, $M_{\rm GUT}/M_{\rm Pl}$ is 
indeed a small number around $10^{-3}-10^{-2}$.
Thus, it is natural to assume that
$M_{\rm GUT}/M_{\rm Pl}$ is small from the effective
field theory point of view in the bottom-up approach, and
then gravity can be decoupled.
In the decoupling limit where $M_{\rm Pl} \rightarrow \infty$
while $M_{\rm GUT}$ remains finite, semi-realistic
$SU(5)$ models and $SO(10)$ models without chiral exotic particles
have been constructed locally. To decouple gravity and 
avoid the bulk matter fields on the observable 
seven-branes, we can show that the observable seven-branes should 
wrap a del Pezzo $n$ surface $dP_n$ with $n \ge 2$ for the internal 
space dimensions (For a review of del Pezzo $n$ surfaces, 
see Appendix A.)~\cite{Beasley:2008dc, Beasley:2008kw}. 
The GUT gauge fields are on the worldvolume of the
observable seven-branes, while the matter and 
Higgs fields are localized on the codimension-one curves 
in $dP_n$. A brand new feature is that the $SU(5)$ and 
$SO(10)$ gauge symmetries can be broken down to the SM and 
$SU(5)\times U(1)$ gauge symmetries,
respectively by turning on $U(1)$ fluxes. Because the $SO(10)$ models 
have not only gauge interaction unification but also fermion unification, 
it seems to us that $SO(10)$ models is more interesting than 
$SU(5)$. In the $SO(10)$ models, to eliminate the zero modes of
the chiral exotic particles,  we must break the $SO(10)$ gauge
symmetry down to the flipped $SU(5)\times U(1)_X$ gauge 
symmetry~\cite{Beasley:2008kw}.
Interestingly, in flipped $SU(5)\times U(1)_X$ 
models~\cite{smbarr, dimitri}, 
we can solve the doublet-triplet splitting problem via 
the missing partner mechanism~\cite{AEHN-0}.

In flipped $SU(5)\times U(1)_X$ models of
$SO(10)$ origin, there are two unification
scales: the $SU(2)_L \times SU(3)_C$ unification scale $M_{23}$
and the $SU(5)\times U(1)_X$ unfication scale $M_U$
where $M_{23}$ is about the usual GUT scale  around
$2\times 10^{16}$ GeV. To solve the little hierarchy problem between  
the GUT scale and string scale $M_{\rm string}$,
we have introduced extra
vector-like particles, and achieved the string-scale
gauge coupling unification in flipped $SU(5)\times U(1)_X$
models~\cite{Lopez:1995cs, Jiang:2006hf}. 
Similarly, for the flipped $SU(5)\times U(1)_X$ models
from F-theory, we can naturally obtain the decoupling scenario 
where $M_{23}/M_U$ or  $M_{23}/M_{\rm Pl}$ can be small
by introducing the additional 
vector-like particles.

In this paper, we briefly review the flipped 
$SU(5)\times U(1)_X$ models with string-scale gauge 
coupling unification~\cite{Lopez:1995cs, Jiang:2006hf}. We also 
review the F-theory model building. To separate
the mass scales $M_{23}$ and $M_U$ and realize the
decoupling scenario,  we introduce sets of 
vector-like particles in complete 
$SU(5)\times U(1)_X$  multiplets, 
whose contributions to the one-loop beta functions of the
  $U(1)_Y$, $SU(2)_L$ and $SU(3)_C$ gauge symmetries,  
$\Delta b_1$,  $\Delta b_2$ and $\Delta b_3$ respectively, 
satisfy $\Delta b_1 < \Delta b_2 = \Delta b_3$. 
To avoid the Landau pole
problem for the gauge couplings, we can only introduce
five sets of vector-like particles around the TeV scale
which could be observed at the LHC. Moreover, we systematically
construct the flipped $SU(5)\times U(1)_X$ models without
bulk vector-like particles: (i) Type I
models only have the vector-like particles around the TeV scale;
(ii) Type II models only have the vector-like particles
at the intermediate scale; (iii) Type III models 
have the vector-like particles around the TeV scale and the high
scale (for definitions, see Section IV). 
For a complete study, we also construct the 
Type I, Type II and Type III models with one pair
and two pairs of bulk vector-like particles on the observable
seven-branes. Interestingly, these vector-like particles 
can couple to the SM singlet fields, and can obtain masses about
from the TeV scale to the GUT scale via Higgs mechanism.
In addition, we study the gauge coupling 
unification in all of our models without bulk vector-like 
particles, and in the Type IA and Type IIA models 
(for definitions, see Sections IV and V) with
bulk vector-like particles. We also study  
 string-scale gauge coupling unification defined
in Eq.~\ref{String-U} in the Type III models,
and the Type IA and Type IIA models with bulk vector-like particles.
We show that the $U(1)_X$ flux contributions to the gauge
couplings preserve the $SU(5)\times U(1)_X$ gauge coupling
unification. We calculate the $SU(3)_C\times SU(2)_L$ unification scales, 
and the $SU(5)\times U(1)_X$ unification scales
and unified couplings. 
Interesetingly, in most of our models,
the high-scale or bulk vector-like particles can be considered
as the string-scale threshold corrections since their
masses are close to the string scale.
We show that the $Z0$ and $Z1$ sets  of vector-like particles 
(for definitions, see Section II) can have masses below the 
$1~{\rm TeV}$ scale, and then they can be produced at the LHC.
Thus, the corresponding
models, which have $Z0$ or $Z1$ sets of vector-like particles 
at about $1~{\rm TeV}$ scale, can be tested at the LHC.

Furthermore, we  discuss the phenomenological consequences
of our models:
(i) We point out that there may exist additional chiral exotic particles
or vector-like particles when we embed the local F-theory GUTs into
the global consistent setup.
(ii) Considering suitable threshold corrections at 
the supersymmetry breaking scale and the $M_{23}$ scale, 
we might have the $Z2$, $Z3$ and $Z4$ sets of vector-like particles 
(for definitions, see Section II) whose masses can be below 
the $1~{\rm TeV}$ scale. Thus, all of our models with
TeV-scale vector-like particles could be tested at the LHC.
(iii) There are Yukawa interactions between the MSSM Higgs
fields and the TeV-scale vector-like particles. With relatively
large Yuakwa couplings which are consistent with the perturbative
unification, we can increase the lightest CP-even Higgs boson mass.
(iv) Considering proton decay $p  \to e^+ \pi^0$ 
via dimension-6 operator from heavy gauge boson exchange
and including the threshold corrections,
we obtain that the proton life time in our models is smaller
than $1\times 10^{35}$ years~\cite{Jiang:2008yf}. 
Thus, our models can definitely 
be tested at the future Hyper-Kamiokande proton decay 
experiment~\cite{Nakamura:2003hk}.
(v) The neutrino masses and mixings can be generated via
the seesaw mechanism, and the baryon asymmetry can be explained
via the leptogenesis. (vi) We can naturally realize the hybrid
inflation in our models, solve the monopole problem, and 
obtain the correct cosmic primodial density 
fluctuations.

This paper is organized as follows: in Section II, we briefly
review the flipped $SU(5)\times U(1)_X$ models.
In Section III, we review the F-theory model buildings,
and discuss the minimal flipped $SU(5)\times U(1)_X$ model.
In Sections IV and V, 
we systematically construct the flipped $SU(5)\times U(1)_X$
models without and with bulk vector-like particles, respectively.
We discuss the phenomenological consequences in Section VI.
Our discussion and conclusions are in Section VI.
In Appendix A, we briefly review the del Pezzo surfaces.
In Appendices B and C, we present the vector-like particle
 curves and the gauge bundle assignments in our models with
one pair and two paris of bulk vector-like particles, respectively.
In Appendix D, we give the one-loop and two-loop 
beta functions for the bulk vector-like particles. 

\section{Flipped $SU(5)\times U(1)_X$ Models}

We first briefly review the minimal flipped
$SU(5)$ model~\cite{smbarr, dimitri, AEHN-0}. 
The gauge group for flipped $SU(5)$ model is
$SU(5)\times U(1)_{X}$, which can be embedded into $SO(10)$ model.
We define the generator $U(1)_{Y'}$ in $SU(5)$ as 
\bea 
T_{\rm U(1)_{Y'}}={\rm diag} \left(-{1\over 3}, -{1\over 3}, -{1\over 3},
 {1\over 2},  {1\over 2} \right).
\label{u1yp}
\eea
The hypercharge is given by
\bea
Q_{Y} = {1\over 5} \left( Q_{X}-Q_{Y'} \right).
\label{ycharge}
\eea

There are three families of the SM fermions 
whose quantum numbers under $SU(5)\times U(1)_{X}$ are
\bea
F_i={\mathbf{(10, 1)}},~ {\bar f}_i={\mathbf{(\bar 5, -3)}},~
{\bar l}_i={\mathbf{(1, 5)}},
\label{smfermions}
\eea
where $i=1, 2, 3$. The SM particle assignments in $F_i$, ${\bar f}_i$ 
and ${\bar l}_i$ are
\bea
F_i=(Q_i, D^c_i, N^c_i),~{\overline f}_i=(U^c_i, L_i),~{\overline l}_i=E^c_i~,~
\label{smparticles}
\eea
where $Q_i$ and $L_i$ are respectively the superfields of the left-handed
quark and lepton doublets, $U^c_i$, $D^c_i$, $E^c_i$ and $N^c_i$ are the
$CP$ conjugated superfields for the right-handed up-type quarks,
down-type quarks, leptons and neutrinos, respectively.
To generate the heavy right-handed neutrino masses, we introduce
three SM singlets $\phi_i$.

To break the GUT and electroweak gauge symmetries, we introduce two pairs
of Higgs representations
\bea
H={\mathbf{(10, 1)}},~{\overline{H}}={\mathbf{({\overline{10}}, -1)}},
~h={\mathbf{(5, -2)}},~{\overline h}={\mathbf{({\bar {5}}, 2)}}.
\label{Higgse1}
\eea
We label the states in the $H$ multiplet by the same symbols as in
the $F$ multiplet, and for ${\overline H}$ we just add ``bar'' above the fields.
Explicitly, the Higgs particles are
\bea
H=(Q_H, D_H^c, N_H^c)~,~
{\overline{H}}= ({\overline{Q}}_{\overline{H}}, {\overline{D}}^c_{\overline{H}}, 
{\overline {N}}^c_{\overline H})~,~\,
\label{Higgse2}
\eea
\bea
h=(D_h, D_h, D_h, H_d)~,~
{\overline h}=({\overline {D}}_{\overline h}, {\overline {D}}_{\overline h},
{\overline {D}}_{\overline h}, H_u)~,~\,
\label{Higgse3}
\eea
where $H_d$ and $H_u$ are one pair of Higgs doublets in the MSSM.
We also add one singlet $\Phi$.

To break the $SU(5)\times U(1)_{X}$ gauge symmetry down to the SM
gauge symmetry, we introduce the following Higgs superpotential at the GUT scale
\bea
{\it W}_{\rm GUT}=\lambda_1 H H h + \lambda_2 {\overline H} {\overline H} {\overline
h} + \Phi ({\overline H} H-M_{\rm H}^2)~.~ 
\label{spgut} 
\eea 
There is only
one F-flat and D-flat direction, which can always be rotated along
the $N^c_H$ and ${\overline {N}}^c_{\overline H}$ directions. So, we obtain that
$<N^c_H>=<{\overline {N}}^c_{\overline H}>=M_{\rm H}$. In addition, the
superfields $H$ and ${\overline H}$ are eaten and acquire large masses via
the supersymmetric Higgs mechanism, except for $D_H^c$ and 
${\overline {D}}^c_{\overline H}$. And the superpotential $ \lambda_1 H H h$ and
$ \lambda_2 {\overline H} {\overline H} {\overline h}$ couple the $D_H^c$ and
${\overline {D}}^c_{\overline H}$ with the $D_h$ and ${\overline {D}}_{\overline h}$,
respectively, to form the massive eigenstates with masses
$2 \lambda_1 <N_H^c>$ and $2 \lambda_2 <{\overline {N}}^c_{\overline H}>$. So, we
naturally have the doublet-triplet splitting due to the missing
partner mechanism~\cite{AEHN-0}. 
Because the triplets in $h$ and ${\overline h}$ only have
small mixing through the $\mu$ term, the Higgsino-exchange mediated
proton decay are negligible, {\it i.e.},
we do not have the dimension-5 proton
decay problem.

The SM fermion masses are from the following
superpotential
\bea 
{ W}_{\rm Yukawa} = y_{ij}^{D}
F_i F_j h + y_{ij}^{U \nu} F_i  {\overline f}_j {\overline
h}+ y_{ij}^{E} {\overline l}_i  {\overline f}_j h + \mu h {\overline h}
+ y_{ij}^{N} \phi_i {\overline H} F_j~,~\,
\label{potgut}
\eea
where $y_{ij}^{D}$, $y_{ij}^{U \nu}$, $y_{ij}^{E}$ and $y_{ij}^{N}$
are Yukawa couplings, and $\mu$ is the bilinear Higgs mass term.

After the $SU(5)\times U(1)_X$ gauge symmetry is broken down to the SM gauge 
symmetry, the above superpotential gives 
\bea 
{ W_{SSM}}&=&
y_{ij}^{D} D^c_i Q_j H_d+ y_{ji}^{U \nu} U^c_i Q_j H_u
+ y_{ij}^{E} E^c_i L_j H_d+  y_{ij}^{U \nu} N^c_i L_j H_u \nnb \\
&& +  \mu H_d H_u+ y_{ij}^{N} 
\langle {\overline {N}}^c_{\overline H} \rangle \phi_i N^c_j
 + \cdots (\textrm{decoupled below $M_{GUT}$}). 
\label{poten1}
\eea

Similar to the flipped $SU(5)\times U(1)_X$ models
with string-scale gauge coupling 
unification~\cite{Lopez:1995cs, Jiang:2006hf},
we introduce vector-like particles which form the complete 
flipped $SU(5)\times U(1)_X$ multiplets.
The quantum numbers for these additional vector-like particles
 under the $SU(5)\times U(1)_X$ gauge symmetry are
\begin{eqnarray}
&& XF ={\mathbf{(10, 1)}}~,~{\overline{XF}}={\mathbf{({\overline{10}}, -1)}}~,~\\
&& Xf={\mathbf{(5, 3)}}~,~{\overline{Xf}}={\mathbf{({\overline{5}}, -3)}}~,~\\
&& Xl={\mathbf{(1, -5)}}~,~{\overline{Xl}}={\mathbf{(1, 5)}}~,~\\
&& Xh={\mathbf{(5, -2)}}~,~{\overline{Xh}}={\mathbf{({\overline{5}}, 2)}}~,~ \\
&& XT ={\mathbf{(10, -4)}}~,~{\overline{XT}}={\mathbf{({\overline{10}}, 4)}}~.~\,
\end{eqnarray}

Moreover,  the particle contents from the decompositions of
$XF$, ${\overline{XF}}$, $Xf$, ${\overline{Xf}}$,
$Xl$, ${\overline{Xl}}$, $Xh$, ${\overline{Xh}}$,
$XF$, and ${\overline{XT}}$, under the SM gauge
symmetry are
\begin{eqnarray}
&& XF = (XQ, XD^c, XN^c)~,~ {\overline{XF}}=(XQ^c, XD, XN)~,~\\
&& Xf=(XU, XL^c)~,~ {\overline{Xf}}= (XU^c, XL)~,~\\
&& Xl= XE~,~ {\overline{Xl}}= XE^c~,~\\
&& Xh=(XD, XL)~,~ {\overline{Xh}}= (XD^c, XL^c)~,~\\
&& XT = (XY, XU^c, XE)~,~ {\overline{XT}}=(XY^c, XU, XE^c)~.~\,
\end{eqnarray}
Under the $SU(3)_C \times SU(2)_L \times U(1)_Y$ gauge
symmetry, the quantum numbers for the extra vector-like 
particles are
\begin{eqnarray}
&& XQ={\mathbf{(3, 2, {1\over 6})}}~,~
XQ^c={\mathbf{({\bar 3}, 2,-{1\over 6})}} ~,~\\
&& XU={\mathbf{({3},1, {2\over 3})}}~,~
XU^c={\mathbf{({\bar 3},  1, -{2\over 3})}}~,~\\
&& XD={\mathbf{({3},1, -{1\over 3})}}~,~
XD^c={\mathbf{({\bar 3},  1, {1\over 3})}}~,~\\
&& XL={\mathbf{({1},  2,-{1\over 2})}}~,~
XL^c={\mathbf{(1,  2, {1\over 2})}}~,~\\
&& XE={\mathbf{({1},  1, {-1})}}~,~
XE^c={\mathbf{({1},  1, {1})}}~,~\\
&& XN={\mathbf{({1},  1, {0})}}~,~
XN^c={\mathbf{({1},  1, {0})}}~,~\\
&& XY={\mathbf{({3}, 2, -{5\over 6})}}~,~
XY^c={\mathbf{({\bar 3}, 2, {5\over 6})}} ~.~\
\end{eqnarray}

To separate the mass scales $M_{23}$ and $M_U$ in our F-theory 
flipped $SU(5)\times U(1)_X$ models,
we need to introduce sets of vector-like particles 
around the TeV scale or intermediate scale whose contributions to the one-loop
beta functions satisfy $\Delta b_1 < \Delta b_2 = \Delta b_3$. 
To avoid the Landau pole problem, we have shown that there are 
only five possible such sets of vector-like
 particles as follows
due to the quantizations of the one-loop beta functions~\cite{Jiang:2006hf}
\begin{eqnarray}
&& Z0: XF+{\overline{XF}}~;~\\
&& Z1: XF+{\overline{XF}}+Xl+{\overline{Xl}} ~;~\\
&&  Z2: XF+{\overline{XF}}+Xf+{\overline{Xf}} ~;~\\
&&  Z3: XF+{\overline{XF}} + Xl+{\overline{Xl}}
+Xh+{\overline{Xh}}  ~;~\\
&&  Z4: XF+{\overline{XF}}+Xh+{\overline{Xh}}~.~\,
\end{eqnarray}
Thus, we will construct the flipped $SU(5)\times U(1)_X$ models with
these sets of vector-like particles around the TeV scale,
and two models respectively with $Z0$ and $Z4$ sets 
at the intermedate scale.

\section{F-Theory Model Building}

We first briefly review the F-theory model 
building~\cite{Vafa:1996xn, Donagi:2008ca,
Beasley:2008dc, Beasley:2008kw, Donagi:2008kj}.
The twelve-dimensional F theory is a convenient way to describe 
Type IIB vacua with varying
axion-dilaton $\tau=a+ie^{-\phi}$. We compactify F-theory on a
Calabi-Yau fourfold, which is elliptically fibered $\pi: Y_4 \to B_3$
with a section $\sigma: B_3 \to Y_4$. The base $B_3$ is the internal
space dimensions in Type IIB string theory, and the complex structure
of the $T^2$ fibre encodes $\tau$ at each point of $B_3$. The SM or GUT
gauge theories are on the worldvolume of the observable
seven-branes that wrap
a complex codimension-one suface in $B_3$. Denoting the complex
coordinate tranverse to these seven-branes in $B_3$ as $z$, we can
write the elliptic fibration in  Weierstrass form
\begin{eqnarray}
 y^2=x^3+f(z)x+g(z)~,~\,
\end{eqnarray}
where $f(z)$  and $g(z)$ are sections of $K_{B_3}^{-4}$ and
$K_{B_3}^{-6}$, respectively. The complex structure of the fibre is 
\begin{eqnarray}
 j(\tau)~=~ {{4(24f)^3}\over {\Delta}}~,~~~
\Delta~=~ 4 f^3 + 27 g^2 ~.~\,
\end{eqnarray}
At the discriminant locus $\{\Delta=0\} \subset B_3$,
the torus $T^2$ degenerates by pinching one of its
cycles and becomes singular. For a generic pinching one-cycle 
 $(p, q)=p\alpha+q\beta$ where $\alpha$ and $\beta$
are one-cylces for the torus $T^2$, we obtain a $(p,q)$ seven-brane in 
the locus where the $(p,q)$ string can end.
The singularity types of the ellitically fibres fall into the 
familiar $ADE$ classifications, and we identify the corresponding 
$ADE$ gauge groups on the seven-brane world-volume. 
This is one of the most important advantages for the F-theory model building: 
the exceptional gauge groups appear rather naturally, which is absent in
perturbative Type II string theory. And then all the SM fermion Yuakwa
couplings in the GUTs can be generated.

We assume that the observable seven-branes with GUT models
on its worldvolume wrap a complex codimension-one 
suface $S$ in $B_3$, and the observable gauge symmetry
is $G_S$. When $h^{1,0}(S)\not=0$, the low energy
spectrum may contain the extra states obtained
by reduction of the bulk supergravity modes of
compactification. So we require that 
$\pi_1(S)$ be a finite group. In order to decouple
gravity and construct models locally, the extension 
of the local metric on $S$ to
a local Calabi-Yau fourfold must have a limit where
the surface $S$ can be shrunk to zero size. This implies
that the anti-canonical bundle on $S$ must be ample. 
Therefore, $S$ is a del Pezzo $n$ surface $dP_n$ with $n \ge 2$
in which $h^{2,0}(S)=0$.
By the way, the Hirzebruch surfaces with degree larger than 2
satisfy $h^{2,0}(S)=0$ but do not define the fully 
consistent decoupled models~\cite{Beasley:2008dc, Beasley:2008kw}.

 To describe the spectrum, we have to study the gauge theory 
of the worldvolume on the seven-branes.  We 
start from the maximal supersymmetric gauge theory on
$\mathbb{R}^{3,1}\times \mathbb{C}^{2}$ and then replace
$\mathbb{C}^{2}$ with the K\"ahler surface $S$. In order to have
four-dimensional ${\cal N}=1$ supersymmetry, the
maximal supersymmetric gauge theory on $\mathbb{R}^{3,1}\times
\mathbb{C}^{2}$ should be twisted. It was shown that there exists a
unique twist preserving ${\cal N}=1$ supersymmetry in four
dimensions, and chiral matters can arise from the bulk $S$ or the
codimension-one curve $\Sigma$ in $S$ which is the intersection
between the observable seven-branes and 
the other seven-brane(s)~\cite{Beasley:2008dc, Beasley:2008kw}.

In order to have the matter fields on $S$,
we consider a non-trivial vector bundle on $S$ with
a structure group $H_S$ which is a subgroup of $G_S$. Then the gauge
group $G_S$ is broken down to $\Gamma_S\times H_S$, and the adjoint
representation ${\rm ad}(G_S)$ of the $G_S$ is decomposed as 
\begin{equation}
{\rm ad}(G_S)\rightarrow
{\rm ad}(\Gamma_S)\bigoplus {\rm ad}(H_S)\bigoplus_j(\tau_j,T_j)~.~\,
\end{equation}
Employing the vanishing theorem of the del Pezzo surfaces,
we obtain the numbers of the generations and anti-generations 
by calculating the zero modes of the Dirac operator on $S$
\begin{eqnarray}
 n_{\tau_j}~=~ -\chi (S, \mathbf{T_j})~,~~~ 
n_{\tau_j^*}~=~ -\chi (S, \mathbf{T_j}^*)~,~\,
\end{eqnarray}
where $\mathbf{T_j}$ is the vector bundle on $S$ whose 
sections transform in the representation $T_j$ of $H_S$,
and $\mathbf{T_j}^*$ is the dual bundle of $\mathbf{T_j}$.
In particular, when the $H_S$ bundle is a line bundle $L$,
we have
\begin{eqnarray}
n_{\tau_j}~=~-\chi (S, L^j)~=~
-\Big[1+\frac{1}{2}\big(\int_{S}c_{1}({L}^{j})c_{1}(S)+
\int_{S}c_{1}({L}^{j})^2\big)\Big]~.~\,
\label{EulerChar}
\end{eqnarray}
In order to preserve supersymmetry, the line bundle $L$ should satisfy
 the BPS equation~\cite{Beasley:2008dc}
\begin{equation}
J_{S}\wedge c_{1}(L)=0,\label{BPS}
\end{equation}
where $J_{S}$ is the K\"ahler form on $S$. Moreover,
the admissible supersymmetric line bundles on del Pezzo surfaces must
satisfy $c_{1}(L)c_{1}(S)=0$, thus,
 $n_{\tau_j}=n_{\tau_j^*}$ and only the vector-like particles
can be obtained. In short, we can not have the chiral matter fields
on the worldvolume of the observable seven-branes.

Interestingly, the chiral superfields can come from the intersections
between the observable seven-branes and the other 
seven-brane(s)~\cite{Beasley:2008dc, Beasley:2008kw}. 
Let us consider a stack of seven-branes with gauge group
$G_{S'}$ that wrap a codimension-one surface $S'$ in $B_3$.
The intersection of $S$ and $S'$ is a codimenion-one curve
(Riemann surface) $\Sigma$ in $S$ and $S'$, 
and the gauge symmetry on $\Sigma$
will be enhanced to $G_{\Sigma}$ 
where $G_{\Sigma}\supset G_{S}\times G_{S'}$.
On this curve, there exist chiral matters from 
the decomposition of the adjoint representation
${\rm ad}G_{\Sigma}$ of $G_{\Sigma}$ as follows
\begin{equation}
{\rm ad}G_{\Sigma}={\rm ad}G_{S}\oplus {\rm ad}G_{S'}\oplus_{k}
({ U}_{k}\otimes { U'}_{k})~.~\,
\end{equation}
Turning on the non-trivial gauge bundles on $S$ and $S'$ respectively
with structure groups $H_S$ and $H_{S'}$, we break the gauge 
group $G_S\times G_{S'}$ down to the commutant subgroup 
$\Gamma_{S}\times\Gamma_{S'}$. Defining 
$\Gamma \equiv \Gamma_{S}\times\Gamma_{S'}$ and 
$H \equiv H_{S}\times H_{S'}$,
we can decompose ${ U}\otimes { U'}$ into the irreducible
representations as follows
\begin{equation}
{ U}\otimes { U'}={\bigoplus}_{k}(r_{k}, {V}_{k}),
\end{equation}
where $r_{k}$ and ${ V}_{k}$ are the representations of $\Gamma$
and $H$, respectively. The light chiral fermions in the
representation $r_{k}$ are determined by the zero modes of the
Dirac operator on $\Sigma$. The net number of chiral superfields
 is given by
\begin{eqnarray}
N_{r_{k}}-N_{r^{*}_{k}}=\chi(\Sigma,K^{1/2}_{\Sigma}\otimes
{\mathbf{V}_{k}}),
\end{eqnarray}
where $K_{\Sigma}$ is the  restriction of
canonical bundle on the curve $\Sigma$, and
$\mathbf{V}_{k}$ is the vector bundle whose sections 
transform in the representation ${ V}_{k}$ of 
the structure group $H$. 

In the F-theory model building, we are interested in the 
models where $G_{S'}$ is $U(1)'$, and
$H_S$ and $H_{S'}$ are respectively $U(1)$
and $U(1)'$. Then the vector bundles on $S$ and $S'$ 
are line bundles $L$ and $L'$. The adjoint representation
${\rm ad}G_{\Sigma}$ of $G_{\Sigma}$ is 
decomposed into a direct sum
of the irreducible representations under the group
$\Gamma_S \times U(1) \times U(1)'$ that can be
denoted as $\mathbf{(r_j, q_j, q'_j)}$
\begin{equation}
{\rm ad}G_{\Sigma}={\rm ad}(\Gamma_S)
\oplus {\rm ad}G_{S'}\oplus_{j}
\mathbf{(r_j, q_j, q_j')}~.~\,
\end{equation}
The numbers of chiral supefields  in the representation 
$\mathbf{(r_j, q_j, q'_j)}$ and their Hermitian conjugates
on the curve $\Sigma$ are given by 
\begin{eqnarray}
N_{\mathbf{(r_j, q_j, q'_j)}} ~=~ h^0 (\Sigma, \mathbf{V}_j) ~,~~~
N_{\mathbf{({\bar r}_j, -q_j, -q'_j)}} ~=~ h^1(\Sigma, \mathbf{V}_j)~,~\,
\end{eqnarray}
where 
\begin{eqnarray}
\mathbf{V}_j~=~ K^{1/2}_{\Sigma} \otimes
{L}_{\Sigma}^{q_{j}}\otimes {L'}_{\Sigma}^{q'_{j}} ~,~\,
\end{eqnarray}
where $K^{1/2}_{\Sigma}$, ${L}_{\Sigma}^{r_{j}}$ and
${L'}_{\Sigma}^{q'_{j}}$ are the restrictions of
canonical bundle $K_S$, line bundles $L$ and $L'$ on the curve
$\Sigma$, respectively. In particular, if the
volume of $S'$ is infinite, $G_{S'}=U(1)'$ is decoupled.
And then the index $\mathbf{q'_j}$ can be ignored.

Using Riemann-Roch theorem, we obtain the net number of 
chiral supefields in the representation $\mathbf{(r_j, q_j, q'_j)}$
\begin{eqnarray}
N_{\mathbf{(r_j, q_j, q'_j)}}-
N_{\mathbf{({\bar r}_j, -q_j, -q'_j)}}~=~ 1-g+{\rm deg}(\mathbf{V}_j) ~,~\,
\end{eqnarray}
where $g$ is the genus of the curve $\Sigma$.

Moreover, we can obtain the Yukawa couplings 
at the triple intersection of
three curves $\Sigma_i$, $\Sigma_j$ and $\Sigma_k$ where
 the gauge group or the singularity type is enhanced further.
To have the triple intersections, the corresponding
homology classes  $[\Sigma_i]$, $[\Sigma_j]$ and $[\Sigma_k]$
of the curves $\Sigma_i$, $\Sigma_j$ and $\Sigma_k$ must satisfy
the following conditions
\begin{eqnarray}
[\Sigma_i] \cdot [\Sigma_j] > 0 ~,~~~
[\Sigma_i] \cdot [\Sigma_k] > 0 ~,~~~
[\Sigma_j] \cdot [\Sigma_k] > 0 ~.~\,
\label{FTYK-Con}
\end{eqnarray}

In this paper, we will construct flipped $SU(5)\times U(1)_X$
models systematically. Thus, we will choose $G_S=SO(10)$ and 
$H_S=U(1)_X$. Under $SU(5)\times U(1)_X$, the $SO(10)$ representations
are decomposed as follows
\begin{eqnarray}
\mathbf{10} &=& \mathbf{(5, -2)} \oplus \mathbf{({\overline{5}}, 2)}~,~ \\
\mathbf{16} &=& \mathbf{(10, 1)} \oplus \mathbf{({\overline{5}}, -3)}
\oplus \mathbf{(1, 5)}~,~\\
\mathbf{45} &=&  \mathbf{(24, 0)} \oplus \mathbf{(1, 0)}
\oplus \mathbf{(10, -4)} \oplus \mathbf{({\overline{10}}, 4)}~.~\,
\end{eqnarray}

Moreover, the Higgs fields $h$ and $\overline{h}$ and
the vector-like particles $Xh$ and $\overline{Xh}$ are on the curves 
where the $SO(10)$ gauge symmetry is enhanced to $SO(12)$. 
Under $SO(10)\times U(1)$, the adjoint representation of $SO(12)$
is decomposed as follows
\begin{eqnarray}
\mathbf{66} &=& \mathbf{(45, 0)} \oplus \mathbf{(1, 0)}
\oplus \mathbf{(10, 2)} \oplus \mathbf{({\overline{10}}, -2)}~.~\,
\end{eqnarray}
All the other fields in our models are on the curves where
the $SO(10)$ gauge symmetry is enhanced to $E_6$. 
Under $SO(10)\times U(1)$, the adjoint representation of $E_6$
is decomposed as follows 
\begin{eqnarray}
\mathbf{78} &=& \mathbf{(45, 0)} \oplus \mathbf{(1, 0)}
\oplus \mathbf{(16, 3)} \oplus \mathbf{({\overline{16}}, -3)}~.~\,
\end{eqnarray}
In addition, the SM fermion Yukawa couplings in our models arise
from the triple intersections where the gauge symmetry is enhanced
to $E_7$.

In this paper, we consider the del Pezzo 8 surface $dP_8$.
In the Section IV, we will choose the line bundle 
$L=\mathcal{O}_{S}(E_{1}-E_{2})^{1/4}$, and construct the
flipped $SU(5)\times U(1)_X$ models without
  bulk vector-like particles $XT_i$ and $\overline{XT}_i$.
Moreover, in the Section V, we will choose the line bundles as
$L=\mathcal{O}_{S}(E_{1}-E_{2}+E_4-E_5)^{1/4}$ and
$L=\mathcal{O}_{S}(E_{1}-E_{2}+E_4-E_5+E_6-E_7)^{1/4}$,
and we construct the flipped $SU(5)\times U(1)_X$ models with
one and two pairs of  
bulk vector-like particles $XT_i$ and $\overline{XT}_i$, respectively.

In our model building, the SM fermion and Higgs curves with homology
classes and the 
gauge bundle assignments for each curve 
in the minimal flipped $SU(5)\times U(1)_X$ model are universal in all
of our models and are given in Table~\ref{tab:mFSU5}. 
In short, all three generations localize on the matter curve $\Sigma_{F}$
which is pinched. Because the homology classes for the SM fermion and
Higgs curves satisfy Eq.~(\ref{FTYK-Con}), the SM fermion Yukawa couplings 
are allowed. There are singlets in the models from the intersections 
of the other seven-branes as well. For simplicity, in the 
following discussions, we will assume the universal supersymmetry 
breaking, and denote the supersymmetry breaking scale as $M_S$.

\begin{table}[htb]
\begin{center}
\begin{tabular}{|c|c|c|c|c|c|}
\hline
${\rm Particles }$ & ${\rm Curve}$ & ${\rm Class}$ & $g_{\Sigma}$ &
$L_{\Sigma}$ & $L_{\Sigma}^{\prime n}$\\\hline
$h$ & $\Sigma_{h}$ & $2H-E_{2}-E_{3}$ & $0$ &
$\mathcal{O}_{\Sigma_{H}^{(d)}}(-1)^{1/4}$ & $\mathcal{O}_{\Sigma_{H}^{(d)}}%
(1)^{1/2}$\\\hline
$\overline{h}$ & $\Sigma_{\overline{h}}$ & $2H-E_{1}-E_{3}$ & $0$ &
$\mathcal{O}_{\Sigma_{H}^{(u)}}(1)^{1/4}$ & $\mathcal{O}_{\Sigma_{H}^{(u)}%
}(-1)^{1/2}$\\\hline
$16_i$ & $\Sigma_{F}\text{ (pinched)}$ & $3H$ & $1$ & $\mathcal{O}%
_{\Sigma_{F}}$ & $\mathcal{O}_{\Sigma_{F}}(3p^{\prime})$\\\hline
$\left(  H+\overline{H}\right)  $ & $\Sigma_{H}{\rm (pinched)}$ &
$3H-E_{1}-E_{2}$ & $1$ & $\mathcal{O}_{\Sigma_{H}}(p_{12})^{1/4}$ &
$\mathcal{O}_{\Sigma_{h}}(p_{12})^{-1/4}$\\\hline
\end{tabular}
\end{center}
\caption{ The SM fermion and Higgs curves
and the gauge bundle assignments for each curve in the minimal flipped 
$SU(5)\times U(1)_X$ model. Here $i=1,~2, ~3$, and $p_{12}=p_1-p_2$.}
\label{tab:mFSU5}
\end{table}

In the following, we will study the gauge couplings in details.
For simplicity, we will neglect the threshold corrections from
the heavy KK modes~\cite{Conlon:2009xf}
since their masses are around the scale $M_U$
and higher in our F-theory flipped $SU(5)\times U(1)_X$ models. 
In addition, the $U(1)_X$ flux will also change the $SU(5)$ and
$U(1)_X$ gauge couplings at the unification 
scale~\cite{Donagi:2008kj, Blumenhagen:2008aw}. From
the particle physics point of view, only the relative changes
between the gauge couplings are physically important while the
total shifts for the gauge couplings are not relevant. For example,
in the F-theory $SU(5)$ models with $U(1)_Y$ flux, only the
$U(1)_Y$ flux contributions to the gauge couplings are relevant,
and the SM gauge coupling relation at the string scale 
is~\cite{Donagi:2008kj, Blumenhagen:2008aw} 
\begin{eqnarray}
\alpha_1^{-1}-\alpha_3^{-1} &=& {3\over 5} (\alpha_2^{-1}-\alpha_3^{-1}) ~.~\,
\end{eqnarray}

Let us consider the flux contributions to the  gauge couplings in our 
F-theory flipped $SU(5)\times U(1)_X$ models. 
For $G=SO(10)$ gauge group, the generators $T^a$ of $SO(10)$ are imaginary
antisymmetric $10 \times 10$ matrices.  In terms of the $2\times 2$
identity matrix $\sigma_0$ and the Pauli matrices $\sigma_i$, they can
be written as tensor products of $2\times 2$ and $5 \times 5$
matrices, $(\sigma_0, \sigma_1, \sigma_3) \otimes A_5$ and $\sigma_2
\otimes S_5$ as a complete set, where $A_5$ and $S_5$ are the $5\times
5$ real anti-symmetric and symmetric matrices~\cite{Huang:2004ui}.  
In particular, the
generator for $U(1)_X$ is $\sigma_2\otimes I_5$ where $I_5$ is the
$5\times 5$ indentiy matrix. Also, the generators for flipped 
$SU(5)\times U(1)_X$  are~\cite{Huang:2004ui}
\begin{eqnarray}
&& \sigma_0 \otimes A_3\,, \quad \sigma_0 \otimes A_2\,, \quad
\sigma_1 \otimes A_X \nonumber \\
&& \sigma_2 \otimes S_3\,, \quad \sigma_2 \otimes S_2\,, \quad
\sigma_3 \otimes A_X ~,~\,
\end{eqnarray}
where $A_3$ and $S_3$ are respectively the diagonal blocks of $A_5$
and $S_5$ that have indices 1, 2, and 3, while the diagonal blocks
$A_2$ and $S_2$ have indices 4 and 5. $A_X$ and $S_X$ are the off
diagonal blocks of $A_5$ and $S_5$.

The flux contributions to the gauge couplings can be computed 
by dimensionally reducing the Chern-Simons action of the
observable seven-branes wrapping on $S$
\begin{eqnarray}
S_{\rm CS} &=& \mu_7 \int_{S\times \mathbb{R}^{3,1}} a \wedge {\rm tr}(F^4) ~.~\,
\end{eqnarray}
In our models, the relevant flux is the $U(1)_X$ flux, which is 
the following 
\begin{eqnarray}
\langle F_{\rm U(1)_X} \rangle &=& {1\over 2} V_{U(1)_X} \sigma_2\otimes I_5   ~.~\,
\end{eqnarray}
Let us noramlize  the $SO(10)$ generators $T^a$ as 
${\rm Tr}(T^aT^b)=2\delta_{ab}$.
Then, we obtain the $U(1)_X$ flux contributions to the $SU(5)$ and $U(1)_X$
gauge couplings at the string scale in our models
\begin{eqnarray}
\Delta \alpha_5^{-1} ~ =~ \Delta \alpha_1^{\prime -1} 
~=~ -{1\over 2} \tau \int_S c_1^2(L^4) ~,~\,
\end{eqnarray}
where $\alpha_1^{\prime}$ is the $U(1)_X$ gauge coupling,
and $\int_S c_1^2(L^4)$ is equal to $-2$, $-4$ and $-6$
for $L=\mathcal{O}_{S}(E_{1}-E_{2})^{1/4}$, 
$L=\mathcal{O}_{S}(E_{1}-E_{2}+E_4-E_5)^{1/4}$ and
$L=\mathcal{O}_{S}(E_{1}-E_{2}+E_4-E_5+E_6-E_7)^{1/4}$, 
respectively.
Because there is no relavant change between the $SU(5)$ and
$U(1)_X$ gauge couplings, the $U(1)_X$ flux contributions
to the gauge couplings are irrelevant from the particle physics
point of view.
In short,   including the $U(1)_X$ flux contributions to
the $SU(5)$ and $U(1)_X$ gauge couplings, we still have
the $SU(5)\times U(1)_X$ gauge coupling unification at the string
scale.

\section{Flipped $SU(5)\times U(1)_X$ Models 
without Bulk Vector-Like Particles}

In this Section, we will take the line bundle 
as $L=\mathcal{O}_{S}(E_{1}-E_{2})^{1/4}$.
Note that $\chi(S, L^4)=0$, we do not have the vector-like particles
$XT_i$ and $\overline{XT}_i$ from the bulk of the observable seven-branes.
In order to separate the mass scales $M_{23}$ and $M_U$ in our F-theory 
flipped $SU(5)\times U(1)_X$ models,
we  introduce sets of vector-like particles 
around the TeV scale, or the intermediate scale, or 
 the TeV scale and high scale.
These vector-like particles can couple to the SM singlet fields
from the intersections of the other seven-branes, and then
obtain masses about from the TeV scale to the GUT scale
by Higgs mechanism because the wave functions for the 
singlet fields  can
be attractive or repulsive and the vacuum expectation values
of the singlet fields are free parameters.

\subsection{Type I Models with TeV-Scale Vector-Like Particles}

In the Type I models, the one-loop contributions to
the beta functions from the sets of vector-like
particles satisfy $\Delta b_2=\Delta b_3$
and $\Delta b_2-\Delta b_1=6/5 $. To avoid the Landau
pole problem for the gauge couplings,
we can only have three  models: Type IA, Type IB and Type IC
models. In the Type IA model,
we introduce $Z1$ set of vector-like
particles. In the Type IB model,
we introduce $Z2$ set of  vector-like
particles. And in the Type IC model,
we introduce $Z3$ set of  vector-like
particles. Also, the curves with homology classes  
for the extra vector-like particles
and  the gauge bundle assignments for each curve in 
Type IA, Type IB and Type IC models are given 
in Table~\ref{tab:T-I-II}. For simplicity,
we assume that the masses for these vector-like particles
are universal, and we denote the universal mass as $M_V$.

\begin{table}[htb]
\begin{center}
\begin{tabular}{|c|c|c|c|c|c|c|}
\hline
${\rm Model}$ &
${\rm Particles }$ & ${\rm Curve}$ & ${\rm Class}$ & $g_{\Sigma}$ &
$L_{\Sigma}$ & $L_{\Sigma}^{\prime n}$\\\hline
Type I \& II &
$\left(  XF+\overline{XF}\right)  $ & $\Sigma_{XF}\text{ (pinched)}$ &
$3H-E_1-E_2-E_{4}$ & $1$ & $\mathcal{O}_{\Sigma_{XF}}(p^4_{12})^{1/4}$ &
$\mathcal{O}_{\Sigma_{XF}}(p^4_{12})^{-1/4}$\\\hline
Type IA & $\left(  Xl+\overline{Xl}\right)  $ & $\Sigma_{Xl}\text{ (pinched)}$ &
$3H-E_{1}-E_{2}-E_5$ & $1$ & $\mathcal{O}_{\Sigma_{Xl}}(p^5_{12})^{1/4}$ &
$\mathcal{O}_{\Sigma_{Xl}}(p^5_{12})^{-5/4}$\\\hline
Type IB &
$\left(  Xf+\overline{Xf}\right)  $ & $\Sigma_{Xf}\text{ (pinched)}$ &
$3H-E_{1}-E_{2}-E_5$ & $1$ & $\mathcal{O}_{\Sigma_{Xf}}(p^5_{12})^{1/4}$ &
$\mathcal{O}_{\Sigma_{Xl}}(p^5_{12})^{3/4}$\\\hline
Type IC & 
$\left(  Xl+\overline{Xl}\right)  $ & $\Sigma_{Xl}\text{ (pinched)}$ &
$3H-E_{1}-E_{2}-E_5$ & $1$ & $\mathcal{O}_{\Sigma_{Xl}}(p^5_{12})^{1/4}$ &
$\mathcal{O}_{\Sigma_{Xl}}(p^5_{12})^{-5/4}$\\
 & $\left(  Xh+\overline{Xh}\right)  $ & $\Sigma_{Xh}\text{ (pinched)}$ &
$3H-E_{1}-E_{2}-E_6$ & $1$ & $\mathcal{O}_{\Sigma_{Xh}}(p^6_{12})^{1/4}$ &
$\mathcal{O}_{\Sigma_{Xh}}(p^6_{12})^{1/2}$\\\hline
Type IIB & $\left(  Xh+\overline{Xh}\right)  $ & $\Sigma_{Xh}\text{ (pinched)}$ &
$3H-E_{1}-E_{2}-E_5$ & $1$ & $\mathcal{O}_{\Sigma_{Xh}}(p^5_{12})^{1/4}$ &
$\mathcal{O}_{\Sigma_{Xh}}(p^5_{12})^{1/2}$\\\hline
\end{tabular}
\end{center}
\caption{ The vector-like particle curves
and the gauge bundle assignments for each curve in Type I and Type II models. 
In particular, we have the vector-like particles ($XF,\overline{XF}$) in all
the Type I and II models. Here, $p_{12}^m=p_1^m-p_2^m$ for $m=4, 5, 6$,
and we denote the corresponding blowing up points as $p_1^m$ or $p_2^m$.}
\label{tab:T-I-II}
\end{table}

\begin{table}[htb]
\begin{center}
\begin{tabular}{|c|c|c|c|c|c|}
\hline
Models &  $M_V$  &  $M_S$  & $M_{23}$   & $g_{\rm U}$ & $M_{\rm U}$  \\
\hline
\hline
Type IA & 200   & 360 & $1.21 \times 10^{16}$   &  1.289 & $6.79
\times 10^{17}$   \\
Type IA & 200   & 1000 & $1.25 \times 10^{16}$   & 1.194 & $6.29
\times 10^{17}$   \\
Type IA & 1000   & 360 & $1.13 \times 10^{16}$   & 1.207  & $1.20
\times 10^{18}$   \\
Type IA & 1000   & 1000 & $1.18 \times 10^{16}$   & 1.143  & $9.33
\times 10^{17}$   \\ 
\hline
\hline
Type IA & $2.0 \times 10^{4}$  & 800 & $1.15 \times 10^{16}$ & 1.051
& $5.54 \times 10^{17}$   \\ 
\hline
Type IB & $2.0 \times 10^{4}$  & 800 & $1.55 \times 10^{16}$ & 1.774
&  $1.04 \times 10^{18}$  \\
\hline
Type IC & $2.0 \times 10^{4}$  & 800 & $1.53 \times 10^{16}$ & 1.790
&  $1.32 \times 10^{18}$  \\
\hline
\end{tabular}
\end{center}
\caption{Mass scales in GeV unit and gauge couplings
in the Type I ${\cal F}-SU(5)$ models for gauge coupling 
unification.  }
\label{tbl:Type-I}
\end{table}

\begin{figure}[htb]
\centering
\includegraphics[width=8cm]{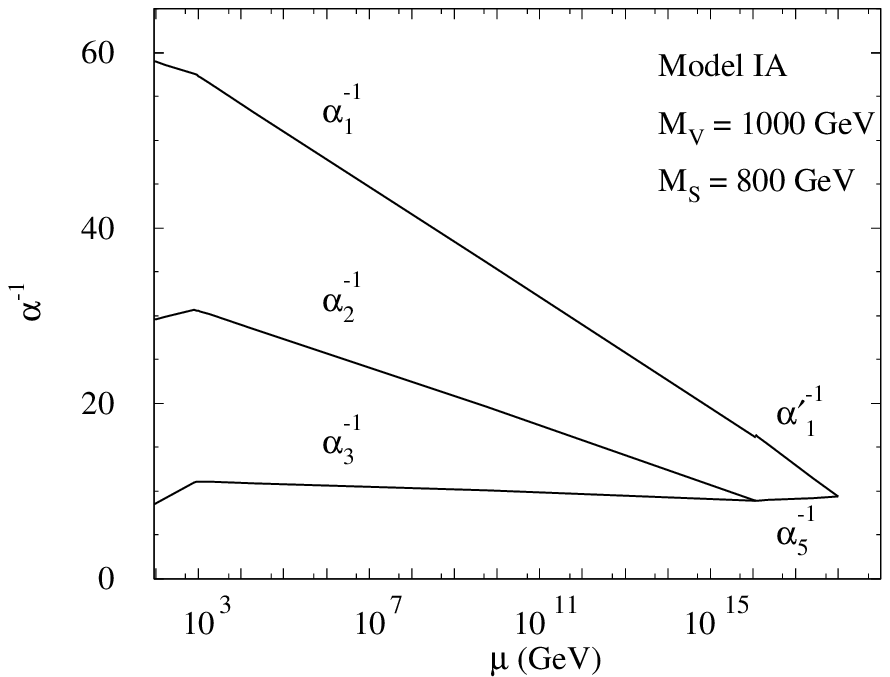}
\includegraphics[width=8cm]{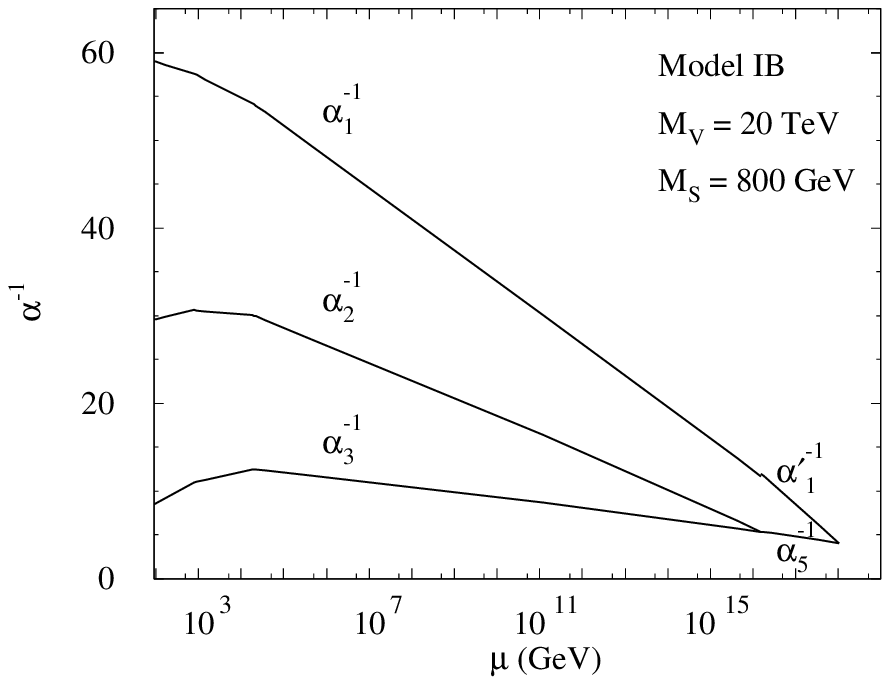}
\caption{Gauge coupling unification in the  
Type IA  model  with $M_V=1 ~{\rm TeV} $ and 
$M_S=800 ~{\rm GeV}$ (left figure), and 
in the Type IB model with 
$M_V=20 ~{\rm TeV}$ and $M_S=800 ~{\rm GeV}$.}
\label{fig:plot12}
\end{figure}

Using the weak-scale data in Ref.~\cite{Amsler:2008zz} 
and the renormalization group equations (RGEs) 
in Ref.~\cite{Jiang:2006hf}, we study the gauge coupling 
unification at the two-loop level. 
In the Type IA models, 
we choose $(M_V, M_S)=(200~{\rm GeV}, ~360~{\rm GeV})$,
$(200~{\rm GeV}, ~1000~{\rm GeV})$, $(1000~{\rm GeV}, ~360~{\rm GeV})$,
 $(1000~{\rm GeV}, ~1000~{\rm GeV})$,
and $(20~{\rm TeV}, ~800~{\rm GeV})$.
We find that $M_{23}$ and $M_U$ are respectively around
 $1.2\times 10^{16}$ GeV and $10^{17-18}$ GeV,
and $g_U$ is about $1.2$. 
In Type IB and Type IC models, to avoid the Landau 
pole problem for gauge couplings, we choose 
$(M_V, M_S)=(20~{\rm TeV}, ~800~{\rm GeV})$.
We present the mass scales $M_{23}$ and $M_U$, 
and the $SU(5)\times U(1)_X$ unified gauge couplings 
$g_U$ in the Type I models in Table~\ref{tbl:Type-I}.
We find that $M_{23}$ and $M_U$ are respectively around
 $1.5\times 10^{16}$ GeV and $10^{18}$ GeV,
and $g_U$ is about $1.8$.   In Fig.~\ref{fig:plot12},
 we plot the gauge coupling unification in the  
Type IA  model  with $M_V=1 ~{\rm TeV} $ and 
$M_S=800 ~{\rm GeV}$, and in the Type IB  
model with $M_V=20 ~{\rm TeV}$
and $M_S=800 ~{\rm GeV} $.
Therefore, only the Type I
models can be tested at the LHC since the universal
mass for $Z_1$ set of vector-like particles can be
below 1 TeV.
We emphasize that the $SU(3)_C\times SU(2)_L$ unified counpling
 $g_{23}$ (very close to $g_U$) 
is stronger than that in the traditional
minimal flipped $SU(5)\times U(1)_X$ models due to the TeV-scale
 vector-like particles, which
will be very important in the proton decay as discussed 
in the Section VI~\cite{Jiang:2008yf}.


\subsection{Type II Models with Intermediate-Scale Vector-Like Particles}

In the Type II models, the one-loop contributions to
the beta functions from the sets of vector-like
particles satisfy $\Delta b_2=\Delta b_3$
and $\Delta b_2-\Delta b_1=12/5$.
We consider two models. In the  Type IIA model, we introduce 
 $Z0$ set of vector-like particles. 
And in the Type IIB model, we introduce
 $Z1$ set of  vector-like particles.
Also, the  curves with homology classes 
for the extra vector-like particles
and the gauge bundle assignments for each curve in Type IIA and 
Type IIB models are given 
in Table~\ref{tab:T-I-II} as well.  For simplicity,
we also assume that the masses for the vector-like particles
are universal, and we denote the universal mass as $M_V$.


\begin{table}[htb]
\begin{center}
\begin{tabular}{|c|c|c|c|c|c|}
\hline
Models &  $M_V$  &  $M_S$  & $M_{23}$   & $g_{\rm U}$ & $M_{\rm U}$  \\
\hline
\hline
Type IIA & $10^{10}$  & 800 & $1.07 \times 10^{16}$ & 0.817 & $6.10
\times 10^{16}$   \\
Type IIA & $10^{11}$  & 800 & $1.06 \times 10^{16}$ & 0.795 & $3.17
\times 10^{17}$   \\
Type IIA & $10^{12}$  & 800 & $1.06 \times 10^{16}$ & 0.774 & $1.67
\times 10^{17}$   \\
\hline
Type IIA & $3.68 \times 10^{10}$ & 800 & $1.08 \times 10^{16}$   &
0.804 & $4.23 \times 10^{17}$   \\
\hline
\hline
Type IIB & $10^{10}$  & 800 & $1.15 \times 10^{16}$ & 0.896 & $6.98
\times 10^{17}$   \\
Type IIB & $10^{11}$  & 800 & $1.12 \times 10^{16}$ & 0.853 & $3.49
\times 10^{17}$   \\
Type IIB & $10^{12}$  & 800 & $1.10 \times 10^{16}$ & 0.816 & $1.78
\times 10^{17}$   \\
\hline
Type IIB & $4.12 \times 10^{10}$ & 800 & $1.14 \times 10^{16}$   &
0.868 & $4.57\times 10^{17}$   \\
\hline
\end{tabular}
\end{center}
\caption{ Mass scales in GeV unit and gauge couplings
in the Type II ${\cal F}-SU(5)$ models
with gauge coupling unification and universal
supersymmetry breaking.  }
\label{tb2:Type-II}
\end{table}

\begin{figure}[htb]
\centering
\includegraphics[width=8cm]{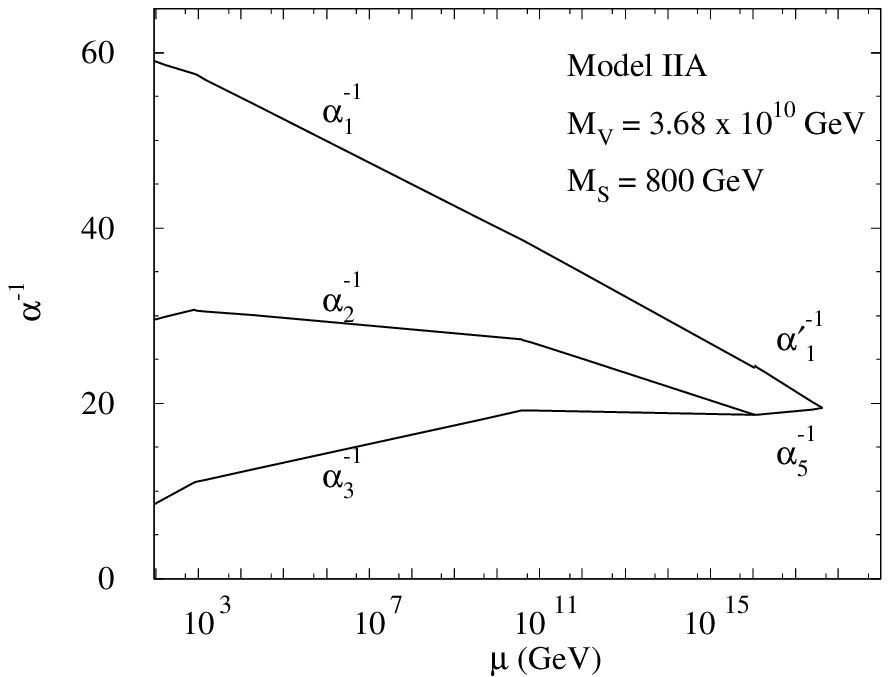}
\includegraphics[width=8cm]{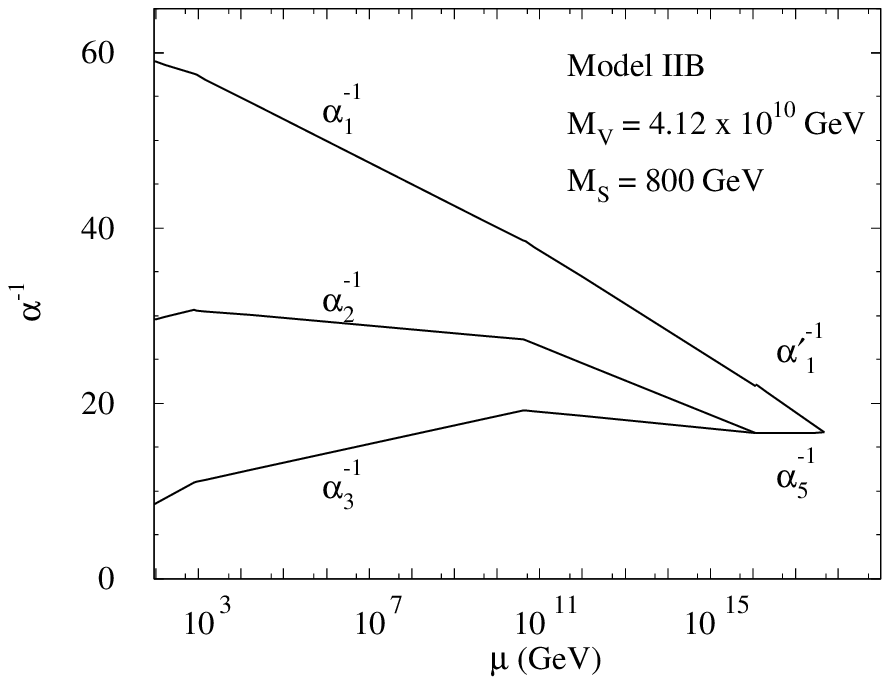}
\caption{String-scale gauge coupling unification 
in the Type IIA model (left figure) and Type IIB model
(right figure) with $M_S=800 ~{\rm GeV} $.}
\label{fig:plot34}
\end{figure}

We study the gauge coupling 
unification in Type II models at the two-loop level. 
Note that $\Delta b_2=\Delta b_3$
and $\Delta b_2-\Delta b_1=12/5$, the $SU(5)\times U(1)_X$
unification scale $M_U$ will be much higher than
the Planck scale $M_{\rm Pl}$ if we put the $Z_0$ or 
$Z_4$ set of vector-like particles around the TeV scale, and
then RGE running must include the supergravity corrections.
Thus, we assume that $M_V$ is at the intermediate scale
 so that we can avoid supergravity corrections to the 
RGE running. Choosing $M_S=800~{\rm GeV}$, and
$M_V=10^{10}~{\rm GeV}$, $10^{11}~{\rm GeV}$
and $10^{12}~{\rm GeV}$,
we present the mass scales $M_{23}$ and $M_U$, 
and the gauge couplings 
$g_U$ in the Type II models in Table~\ref{tb2:Type-II}.
To achieve the string-scale gauge coupling unification
defined in Eq. \ref{String-U},
with $M_S=800~{\rm GeV}$, we obtain that $M_V$ is equal
to $3.68 \times 10^{10}~{\rm GeV}$ in Type IIA model,
and equal to $4.12 \times 10^{10}~{\rm GeV}$
in Type IIB model. We present 
the corresponding mass scales and 
gauge couplings in Table~\ref{tb2:Type-II} as well. 
Moreover, we plot the string-scale gauge coupling 
unification in the Type IIA and Type IIB models in
Fig.~\ref{fig:plot34}. In short,
we find that $M_{23}$ and $M_U$ are respectively around
 $1.1\times 10^{16}$ GeV and $10^{17}$ GeV,
and $g_U$ is about $0.8$. Unfortunately,
Type II models can not be tested at the LHC since
the additional vector-like particles are at the intermediate
scale.




\subsection{Type III  Models with the TeV-Scale and
 High-Scale Vector-Like Particles}

In the Type III models, in addition to the vector-like 
particles around the TeV scale, 
we introduce the high-scale vector-like particles as well.
For simplicity, we also assume that the masses of the high-scale
 vector-like particles are universal, and we denote
their universal mass as $M_{V'}$. In almost all of
the Type III models, $M_V'$ is around the GUT scale 
or higher, so, the high-scale vector-like particles
can be considered as the string-scale threshhold corrections.

We consider four kinds of models. In the Type IIIA models, 
we introduce the following additional vector-like particles:
\begin{eqnarray}
 Z5: XF+{\overline{XF}} +3\times (Xl_i+{\overline{Xl}}_i)~,~\,
\end{eqnarray}
where $i=1,~2,~3$.
In the Type IIIA1 model, we assume that the vector-like 
particles ($XF$, $\overline{XF}$) have masses around the 
TeV scale, while the vector-like particles ($Xl_i$, $\overline{Xl}_i$)
with $i=1,~2,~3$ have masses at the high scale.
In the Type IIIA2 model,
we assume that the vector-like particles ($XF$, $\overline{XF}$) and
($Xl_1$, $\overline{Xl}_1$) have masses around the TeV scale, while
the vector-like particles ($Xl_j$, $\overline{Xl}_j$)
with $j=2,~3$ have masses at the high scale.

In the Type IIIB models, we introduce 
the following extra vector-like
particles:
\begin{eqnarray}
 Z6: XF+{\overline{XF}} +4\times (Xl_k+{\overline{Xl}}_k)~,~\,
\end{eqnarray}
where $k=1,~2,~3,~4$.
In the Type IIIB1 model, we assume that the vector-like particles 
($XF$, $\overline{XF}$) have masses around the TeV scale, while
the vector-like particles ($Xl_k$, $\overline{Xl}_k$)
with $k=1,~2, ~3,~4$ have masses at the high scale. 
In the Type IIIB2 model, we assume that the vector-like particles 
($XF$, $\overline{XF}$) and ($Xl_4$, $\overline{Xl}_4$)
have masses around the TeV scale, while
the vector-like particles ($Xl_i$, $\overline{Xl}_i$)
with $i=1,~2, ~3$ have masses at the high scale.

In the Type IIIC models, we introduce 
the following additional vector-like
particles:
\begin{eqnarray}
 Z7: XF+{\overline{XF}} +  3\times (Xl_i+{\overline{Xl}}_i) + 
Xf+{\overline{Xf}}~,~\,
\end{eqnarray}
where $i=1,~2,~3$.
In the Type IIIC1 model, we assume that  the vector-like 
particles ($XF$, $\overline{XF}$) 
have masses around the TeV scale, while
the vector-like particles ($Xl_i$, $\overline{Xl}_i$) 
with $i=1,~2,~3$, and ($Xf$, $\overline{Xf}$) have
masses at the high scale.
In the Type IIIC2 model, we assume that  the vector-like 
particles ($XF$, $\overline{XF}$) and ($Xl_1$, $\overline{Xl}_1$) 
have masses around the TeV scale, while
the vector-like particles ($Xl_j$, $\overline{Xl}_j$) 
with $j=2,~3$, and ($Xf$, $\overline{Xf}$) have
masses at the high scale.
In the Type IIIC3 model, we assume that the vector-like 
particles ($XF$, $\overline{XF}$) and ($Xf$, $\overline{Xf}$) 
have masses around the TeV scale, while
the vector-like particles ($Xl_i$, $\overline{Xl}_i$), 
with $i=1,~2,~3$ have masses at the high scale.

In the  Type IIID models,
we introduce the following additional vector-like
particles:
\begin{eqnarray}
 Z8: XF+{\overline{XF}} + 3\times (Xl_i+{\overline{Xl}}_i) 
+  Xh+{\overline{Xh}}~,~\,
\end{eqnarray}
where $i=1,~2,~3$.
In the Type IIID1 model, we assume that the vector-like 
particles ($XF$, $\overline{XF}$)
 have masses around the TeV scale, while
the vector-like particles ($Xl_i$, $\overline{Xl}_i$)
with $i=1,~2,~3$,  and ($Xh$, $\overline{Xh}$)  have
masses at the high scale.
In the Type IIID2 model,
we assume that the vector-like particles ($XF$, $\overline{XF}$)
and ($Xl_1$, $\overline{Xl}_1$) 
 have masses around the TeV scale, while
the vector-like particles ($Xl_j$, $\overline{Xl}_j$)
with $j=2,~3$, and ($Xh$, $\overline{Xh}$) have
masses at the high scale.
In the Type IIID3 model,
we assume that the vector-like particles ($XF$, $\overline{XF}$) 
 and ($Xh$, $\overline{Xh}$)
 have masses around the TeV scale, while
the vector-like particles ($Xl_i$, $\overline{Xl}_i$)
with $i=1,~2,~3$ have
masses at the high scale.
In the Type IIID4 model,
we assume that the vector-like particles ($XF$, $\overline{XF}$), 
($Xl_1$, $\overline{Xl}_1$) and ($Xh$, $\overline{Xh}$)
 have masses around the TeV scale, while
the vector-like particles ($Xl_i$, $\overline{Xl}_i$)
with $i=2,~3$ have
masses at the high scale.


\begin{table}[htb]
\begin{center}
\begin{tabular}{|c|c|c|c|c|c|c|c|}
\hline
Models &${\rm Particles}$ & ${\rm Curve}$ & ${\rm Class}$ & $g_{\Sigma}$ &
$L_{\Sigma}$ & $L_{\Sigma}^{\prime n}$\\\hline
Type III & $\left(  XF+\overline{XF}\right)  $ & 
$\Sigma_{XF}\text{ (pinched)}$ & $ 3H-E_1-E_2-E_4 $ & $1$ & 
$\mathcal{O}_{\Sigma_{XF}}(p_{12}^4)^{1/4}$ & $\mathcal{O}_{\Sigma_{XF}}
(p_{12}^4)^{-1/4}$ \\
& $\left(  Xl_i+\overline{Xl}_i\right)  $ & $\Sigma_{Xl_i}\text{ (pinched)}$ &
$3H-E_{1}-E_{2}-E_j$ & $1$ & $\mathcal{O}_{\Sigma_{Xf}}(p^j_{12})^{1/4}$ &
$\mathcal{O}_{\Sigma_{Xf}}(p^j_{12})^{-5/4}$\\\hline
Type IIIA &
$\left(  Xl+\overline{Xl}\right)  $ & $\Sigma_{Xl}\text{ (pinched)}$ &
$3H-E_{1}-E_{2}-E_5$ & $1$ & $\mathcal{O}_{\Sigma_{Xl}}(p^5_{12})^{1/4}$ &
$\mathcal{O}_{\Sigma_{Xf}}(p^5_{12})^{-5/4}$\\\hline
Type IIIB &
$\left(  Xf+\overline{Xf}\right)  $ & $\Sigma_{Xf}\text{ (pinched)}$ &
$3H-E_{1}-E_{2}-E_5$ & $1$ & $\mathcal{O}_{\Sigma_{Xf}}(p^5_{12})^{1/4}$ &
$\mathcal{O}_{\Sigma_{Xf}}(p^5_{12})^{3/4}$\\\hline
Type IIID &
$\left(  Xh+\overline{Xh}\right)  $ & $\Sigma_{Xh}\text{ (pinched)}$ &
$3H-E_{1}-E_{2}-E_5$ & $1$ & $\mathcal{O}_{\Sigma_{Xh}}(p^5_{12})^{1/4}$ &
$\mathcal{O}_{\Sigma_{Xh}}(p^5_{12})^{1/2}$\\\hline
\end{tabular}
\end{center}
\caption{ The vector-like particle curves
and the gauge bundle assignments for each curve in Type III models. 
In particular,  we have the vector-like particles ($XF,\overline{XF}$)
and ($Xl_i$, ${\overline{Xl}}_i$) with $i=1,~2,~3$ in all
the Type III models. Here, $j=i+5$, and $p_{12}^m=p_1^m-p_2^m$ 
for $m=4,~5,~6,~7,~8$. And we denote the corresponding blowing up 
points as $p_1^m$ or $p_2^m$.}
\label{tab:T-III}
\end{table}


Moreover, the  curves with homology classes 
for the additional vector-like particles
and the gauge bundle assignments for each curve in 
Type III models are given 
in Table~\ref{tab:T-III}. We also present 
the complete additional vector-like particles at the scales 
$M_V$ and $M_{V'}$ in Type III models
in the Table \ref{tb3:Type-III}.
In short, at the TeV scale, we have $Z0$ set of 
vector-like particles in the Type IIIX1 models
where X=A, B, C, D; we have $Z1$ set of vector-like 
particles in  the Type IIIX2 models; we
have $Z2$ set of vector-like particles in the 
Type IIIC3 model; we have $Z3$ set
of vector-like particles in  the Type IIID4 model;
and we have $Z4$ set
of vector-like particles in  the Type IIID3 model.



\begin{table}[htb]
\begin{center}
\begin{tabular}{|c|c|c|}
\hline
Models &  Particles at $M_V$    & Particles at  $M_{V'}$   \\
\hline
\hline
Type IIIA1 &  ($XF$, $\overline{XF}$) & ($Xl_i$, $\overline{Xl}_i$) for $i=1,~2,~3$ \\
\hline
Type IIIA2 &  ($XF$, $\overline{XF}$), 
($Xl_1$, $\overline{Xl}_1$) & ($Xl_j$, $\overline{Xl}_j$)
for $j=2,~3$ \\
\hline 
\hline
Type IIIB1 &  ($XF$, $\overline{XF}$) &  ($Xl_k$, $\overline{Xl}_k$)
for $k=1,~2, ~3,~4$ \\
\hline
Type IIIB2 &   ($XF$, $\overline{XF}$), ($Xl_4$, $\overline{Xl}_4$)
& ($Xl_i$, $\overline{Xl}_i$) for $i=1,~2, ~3$ \\
\hline
\hline
Type IIIC1 &  ($XF$, $\overline{XF}$) &
 ($Xl_i$, $\overline{Xl}_i$) 
for $i=1,~2,~3$, ($Xf$, $\overline{Xf}$) \\
\hline
Type IIIC2 & ($XF$, $\overline{XF}$), ($Xl_1$, $\overline{Xl}_1$) 
& ($Xl_j$, $\overline{Xl}_j$) for $j=2,~3$, ($Xf$, $\overline{Xf}$) \\
\hline
Type IIIC3 &  ($XF$, $\overline{XF}$), ($Xf$, $\overline{Xf}$) 
& ($Xl_i$, $\overline{Xl}_i$) for $i=1,~2,~3$ \\
\hline
\hline
Type IIID1 & ($XF$, $\overline{XF}$) 
& ($Xl_i$, $\overline{Xl}_i$) for $i=1,~2,~3$, ($Xh$, $\overline{Xh}$)   \\
\hline
Type IIID2 & ($XF$, $\overline{XF}$), ($Xl_1$, $\overline{Xl}_1$)
&   ($Xl_j$, $\overline{Xl}_j$)
for $j=2,~3$, ($Xh$, $\overline{Xh}$)   \\
\hline
Type IIID3 &  ($XF$, $\overline{XF}$), ($Xh$, $\overline{Xh}$) 
&   ($Xl_i$, $\overline{Xl}_i$) for $i=1,~2,~3$  \\
\hline
Type IIID4 &  ($XF$, $\overline{XF}$), 
($Xl_1$, $\overline{Xl}_1$),  ($Xh$, $\overline{Xh}$)
&  ($Xl_j$, $\overline{Xl}_j$) for $j=2,~3$ \\
\hline
\end{tabular}
\end{center}
\caption{The additional vector-like particles at the scales $M_V$ and $M_{V'}$,
where  $i=1,~2,~3$, $j=2,~3$, $k=1,~2, ~3,~4$. }
\label{tb3:Type-III}
\end{table}



\begin{table}[htb]
\begin{center}
\begin{tabular}{|c|c|c|c|c|c|c|}
\hline
Models &  $M_V$  &  $M_S$  & $M_{23}$  &  $M_{V'}$ & $g_{\rm U}$ 
& $M_{\rm U}$  \\
\hline
\hline
Type IIIA1 & $1000$   & 800 & $1.17 \times 10^{16}$  & $1\times
10^{16}$ & 1.142 & $7.22 \times 10^{18}$   \\
Type IIIA2 & $1000$   & 800 & $1.17 \times 10^{16}$  & $1\times
10^{16}$ & 1.161 & $4.05\times 10^{17}$   \\
\hline
Type IIIB1 & $1000$   & 800 & $1.17 \times 10^{16}$  & $1\times
10^{16}$ & 1.145 & $4.04 \times 10^{18}$   \\
Type IIIB2 & $1000$   & 800 & $1.17 \times 10^{16}$  & $1\times
10^{16}$ & 1.163 & $2.92 \times 10^{17}$   \\
\hline
Type IIIC1 & $1000$   & 800 & $1.17 \times 10^{16}$  & $1\times
10^{16}$ &1.226  & $4.11 \times 10^{18}$   \\
Type IIIC2 & $1000$   & 800 & $1.18 \times 10^{16}$  & $1\times
10^{16}$ & 1.207 & $2.93 \times 10^{17}$   \\
Type IIIC3 & $5000$   & 800 & $1.72 \times 10^{16}$  & $1\times
10^{16}$ &2.470  & $4.41 \times 10^{17}$   \\
\hline
Type IIID1 & $1000$   & 800 & $1.17 \times 10^{16}$  & $1\times
10^{16}$ & 1.231 & $7.77 \times 10^{18}$   \\
Type IIID2 & $1000$   & 800 & $1.17 \times 10^{16}$  & $1\times
10^{16}$ & 1.209 & $4.19 \times 10^{17}$   \\
Type IIID3 & $5.0\times 10^4$   & 800 & $1.50 \times 10^{16}$  &
$1\times 10^{16}$ & 1.613 &  $5.19 \times 10^{18}$  \\
Type IIID4 & $5000$   & 800 & $1.69 \times 10^{16}$  & $1\times
10^{16}$ &3.390 &   $9.97\times 10^{17}$  \\
\hline
\end{tabular}
\end{center}
\caption{Mass scales in GeV unit and gauge couplings
 in the Type III ${\cal F}-SU(5)$ models
with  gauge coupling unification and universal
supersymmetry breaking.   }
\label{tb3:Type-IIIA}
\end{table}



\begin{table}[htb]
\begin{center}
\begin{tabular}{|c|c|c|c|c|c|c|}
\hline
Models &  $M_V$  &  $M_S$  & $M_{23}$  &  $M_{V'}$ & $g_{\rm string}$ 
& $M_{\rm string}$  \\
\hline
\hline
Type IIIA1 & $1000$   & 800 & $1.13\times10^{16}$  &
$2.04\times10^{12}$ &1.161  & $6.12\times 10^{17}$   \\
Type IIIA2 & $1000$   & 800 & $1.18\times 10^{16}$  & $8.32 \times
10^{16}$ & 1.158 & $6.10 \times 10^{17}$   \\
\hline
Type IIIB1 & $1000$   & 800 & $1.13 \times 10^{16}$  & $4.79 \times
10^{12}$ &1.161  & $6.12 \times 10^{17}$   \\
Type IIIB2 & $1000$   & 800 & $1.18 \times 10^{16}$  & $1.62 \times
10^{17}$ & 1.158 & $6.10 \times 10^{17}$   \\
\hline
Type IIIC1 & $1000$   & 800 & $1.20 \times 10^{16}$  & $5.67 \times
10^{12}$ & 1.302 & $6.86 \times 10^{17}$   \\
Type IIIC2 & $1000$   & 800 & $1.18 \times 10^{16}$  & $1.70 \times
10^{17}$ & 1.174 & $6.18 \times 10^{17}$   \\
Type IIIC3 & $1 \times 10^{4}$   & 800 & $1.65 \times 10^{16}$  &
$5.64 \times 10^{17}$ & 2.087  
& $1.10\times 10^{18}$  \\
\hline
Type IIID1 & $1000$   & 800 & $1.24 \times 10^{16}$  & $1.92 \times
10^{12}$ & 1.375 & $7.25 \times 10^{17}$   \\
Type IIID2 & $1000$   & 800 & $1.18 \times 10^{16}$  & $8.55 \times
10^{16}$ & 1.182 & $6.23 \times 10^{17}$   \\
Type IIID3 & $5 \times 10^{4}$   & 800 & $1.50 \times 10^{16}$  &
$2.38 \times 10^{13}$ & 1.533  &
 $8.08 \times 10^{17}$  \\
Type IIID4 & $1 \times 10^{4}$   & 800 & $1.62 \times 10^{16}$  &
$1.46 \times 10^{17}$ & 2.074 & 
 $1.09 \times 10^{18}$    \\
\hline
\end{tabular}
\end{center}
\caption{Mass scales in GeV unit and gauge couplings
 in the Type III ${\cal F}-SU(5)$ models
with string-scale gauge coupling unification and universal
supersymmetry breaking. }
\label{tb3:Type-IIIB}
\end{table}



Furthermore, first, we  study the gauge coupling 
unification in Type III models at the two-loop level. 
We choose $M_S=800~{\rm GeV}$ and 
$M_{V'}=1\times 10^{16}~{\rm GeV}$. For the Type IIIX1
and Type IIIX2 models, we choose
$M_V=1~{\rm TeV}$. To avoid the Landau pole problem
for gauge couplings, we choose $M_V=5~{\rm TeV}$ in
the Type IIIC3 and Type IIID4 models, and choose
$M_V=50~{\rm TeV}$ in the Type IIID3 model.
We present the mass scales $M_{23}$ and $M_U$, 
and the gauge couplings 
$g_U$ in the Type III models in Table~\ref{tb3:Type-IIIA}.
Moreover, we find that  $M_{23}$ and $M_U$ are respectively 
around $1.4\times 10^{16}$ GeV and $10^{17-18}$ GeV.
Also, $g_U$ is about $1.2$ in the Type IIIX1
and Type IIIX2 models,  1.6 in the Type IIID3 model,
2.47 in the Type IIIC3 model, and 3.39 in the Type IIID4 model.
Thus, the unified couplings in the Type IIIC3 and
Type IIID4 models are  strong.

\begin{figure}[htb]
\centering
\includegraphics[width=8cm]{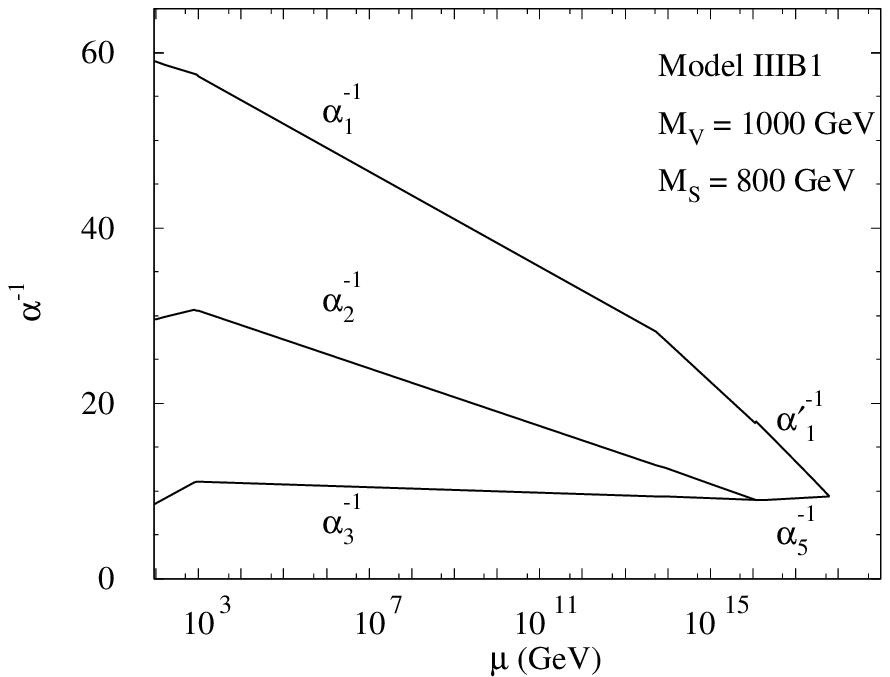}
\includegraphics[width=8cm]{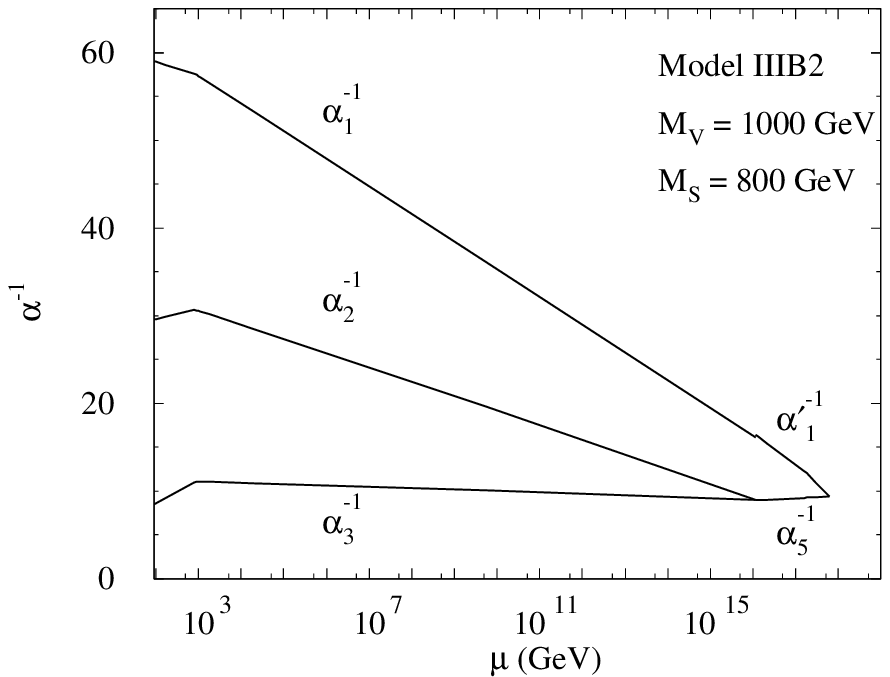}
\caption{String-scale gauge coupling 
unification  in the Type IIIB1 model (left figure) 
and Type IIIB2 model (right figure) with 
$M_V=1 ~{\rm TeV} $ and $M_S=800 ~{\rm GeV} $.}
\label{fig:plot56}
\end{figure}

Second, we study the string-scale gauge coupling 
unification in Type III models at the two-loop level,
and the mass scales $M_{V'}$ are determined from
the condition for string-scale gauge coupling unification
given in Eq.~\ref{String-U}.  
We also choose $M_S=800~{\rm GeV}$. For the Type IIIX1
and Type IIIX2 models, we choose
$M_V=1~{\rm TeV}$. To avoid the Landau pole problem
for gauge couplings, we choose $M_V=10~{\rm TeV}$ in
 the Type IIIC3  and Type IIID4 models,
and  $M_V=50~{\rm TeV}$ in  the Type IIID3 model.
We present the mass scales $M_{23}$ and $M_{\rm string}$, 
and the gauge couplings 
$g_{\rm string}$ in the Type III models in Table~\ref{tb3:Type-IIIB}.
In Fig.~\ref{fig:plot56}, we plot the string-scale gauge coupling 
unification in the Type IIIB1 and Type IIIB2 models
with  $M_V=1 ~{\rm TeV} $ and $M_S=800 ~{\rm GeV} $.
Moreover, we find that  
$M_{23}$ and $M_{\rm string}$ are respectively 
around $1.4\times 10^{16}$ GeV and $10^{17-18}$ GeV.
Also, $g_{\rm string}$ is about $1.2$ in  the Type IIIX1
and Type IIIX2 models, 
 about 1.5 in the Type IIID3 model, and about
2.1 in  the Type IIIC3 and Type IIID4 models.
In addition, in the Type IIIX2, Type IIIC3 and
Type IIID4 models, the high-scale vector-like particles
can be considered as string-scale threshold corrections
since their masses are about $10^{17}$ GeV.
While in the Type IIIX1 and  Type IIID3 models, the masses for 
the vector-like particles are at the intermediate scale 
$10^{12-13}$ GeV.


\section{Flipped $SU(5)\times U(1)_X$ Models 
with Bulk Vector-Like Particles}

In all the above flipped $SU(5)\times U(1)_X$
models, we can introduce the bulk
vector-like particles  $XT_i$ and $\overline{XT}_i$ 
on the observable seven-branes as well.
If we choose the line bundle 
$L=\mathcal{O}_{S}(E_{1}-E_{2}+E_4-E_5)^{1/4}$, we have
one pair of the bulk
 vector-like particles $XT_1$ and $\overline{XT}_1$
on the surface $S$ since $\chi(S, L^4)$ is equal to $-1$. And
if we choose the line bundle 
$L=\mathcal{O}_{S}(E_{1}-E_{2}+E_4-E_5+E_6-E_7)^{1/4}$,
we have two pairs of the bulk
 vector-like particles $XT_i$ and $\overline{XT}_i$
on the surface $S$ since $\chi(S, L^4)$ is equal to $-2$.
For the Type I, Type II and Type III models with
one pair  and two pairs of the bulk 
vector-like particles $XT_i$ and $\overline{XT}_i$,
we present the curves with homology classes for the
vector-like particles, and the gauge bundle
assignments for each curve in Appendices B and C,
respectively. Moreover, the vector-like particles $XT_i$ 
and $\overline{XT}_i$ can obtain
masses via instanton effects. Also, they can couple to the singlets 
from the intersections of the other seven-branes, and then obtain
masses from Higgs mechanism. Thus, the vector-like particles $XT_i$ 
and $\overline{XT}_i$ can have masses $M_{V'}$ close to the string
scale (or intermediate scale) and can be considered as the
string-scale (or intermediate scale) threshold corrections.

To avoid the Landau pole problem for the
gauge couplings, we have shown that only the $Z0$ and $Z1$
sets of vector-like particles can be below
1 TeV, which can be tested at the LHC. Thus, in this 
Section, we will concentrate on the Type IA and Type IIA models
with bulk vector-like particles $XT_i$ and $\overline{XT}_i$.
In the Type IA1 and Type IIA1 models, we introduce one pair of
vector-like particles $XT_1$ and $\overline{XT}_1$.
Also, in the Type IA2 and Type IIA2 models, we introduce 
two pairs of vector-like particles $XT_i$ and $\overline{XT}_i$
with $i=1,~2$. The particle contents of these models  are given in
Table~\ref{A-IIIA12}.

\begin{table}[htb]
\begin{center}
\begin{tabular}{|c|c|c|}
\hline
Models &  Particles at $M_V$    & Particles at  $M_{V'}$   \\
\hline
\hline
~Type IA1 ~& ~ ($XF$, $\overline{XF}$), ($Xl$, $\overline{Xl}$)~
& ~($XT_1$,  $\overline{XT}_1$) ~
  \\
\hline
~Type IA2 ~& ~($XF$, $\overline{XF}$), ($Xl$, $\overline{Xl}$)~
& ~($XT_i$,  $\overline{XT}_i$) for $i=1,~2$ ~\\
\hline 
\hline
~Type IIA1~ &  ($XF$, $\overline{XF}$)
& ($XT_1$,  $\overline{XT}_1$) 
  \\
\hline
~Type IIA2~ & ($XF$, $\overline{XF}$)
& ($XT_i$,  $\overline{XT}_i$) for $i=1,~2$ \\
\hline
\end{tabular}
\end{center}
\caption{The particle contents in the Type IA1, Type IA2, Type IIA1 and
Type IIA2 models. }
\label{A-IIIA12}
\end{table}


\begin{table}[htb]
\begin{center}
\begin{tabular}{|c|c|c|c|c|c|c|}
\hline
Models &  $M_V$  &  $M_S$  & $M_{23}$  &  $M_{V'}$ & $g_{\rm U}$ &
$M_{\rm U}$  \\
\hline
\hline
Type IA1 & 800  & 800 & $1.18 \times 10^{16}$  & $1\times 10^{16}$ &
1.322 & $2.16 \times 10^{17}$   \\
Type IA2 & 800   & 800 & $1.18 \times 10^{16}$  & $1\times 10^{16}$ &
1.428 & $9.85 \times 10^{16}$   \\
Type IIA1 & 800  & 800 & $1.18 \times 10^{16}$  & $1\times 10^{16}$ &
1.527 & $3.87 \times 10^{18}$   \\
Type IIA2 &  800  & 800 & $1.18 \times 10^{16}$  & $1\times 10^{16}$
&
1.996 & $7.53 \times 10^{17}$   \\
\hline
\end{tabular}
\end{center}
\caption{Mass scales in GeV unit and gauge couplings
in Type IA1, Type IA2, Type IIA1 and
Type IIA2 models with gauge coupling unification and universal
supersymmetry breaking.  }
\label{B-IIIA12}
\end{table}

\begin{table}[htb]
\begin{center}
\begin{tabular}{|c|c|c|c|c|c|c|}
\hline
Models &  $M_V$  &  $M_S$  & $M_{23}$  &  $M_{V'}$ & $g_{\rm string}$
& $M_{\rm string}$  \\
\hline
\hline
Type IA1 & 800   & 800 & $1.18 \times 10^{16}$  & $2.46 \times
10^{17}$ & 1.205 & $6.35 \times 10^{17}$   \\
Type IA2 & 800   & 800 & $1.18 \times 10^{16}$  & $3.95 \times
10^{17}$ & 1.205 & $6.35 \times 10^{17}$   \\
Type IIA1 & 800   & 800 & $1.20 \times 10^{16}$  & $2.04 \times
10^{14}$ & 2.020 & $1.06 \times 10^{18}$   \\
\hline
\hline
Type IIA2 & 200   & 360 & $1.21 \times 10^{16}$  & $4.42 \times
10^{16}$ & 4.288 & $2.26 \times 10^{18}$   \\
Type IIA2 & 200   & 1000 & $1.25 \times 10^{16}$  & $1.70 \times
10^{16}$ & 2.153 & $1.13 \times 10^{18}$   \\
Type IIA2 & 1000   & 360 & $1.17 \times 10^{16}$  & $1.91 \times
10^{16}$ & 2.142 & $1.13 \times 10^{18}$   \\
Type IIA2 & 1000   & 1000 & $1.18 \times 10^{16}$  & $1.65 \times
10^{16}$ & 1.862 & $9.81 \times 10^{17}$   \\
\hline
\end{tabular}
\end{center}
\caption{Mass scales in GeV unit and gauge couplings 
in the Type IA1, Type IA2, Type IIA1 and
Type IIA2 models with
string-scale gauge coupling unification and universal
supersymmetry breaking.  }
\label{C-IIIA12}
\end{table}


We give the one-loop and two-loop beta functions for 
$XT_i$ and $\overline{XT}_i$ in the supersymmetric Standard Model and 
in the flipped $SU(5)\times U(1)_X$
model in the Appendix D.
First, we  study the gauge coupling 
unification  at the two-loop level. 
Choosing $M_V=800~{\rm GeV}$, $M_S=800~{\rm GeV}$,
 and $M_{V'}=1\times 10^{16}~{\rm GeV}$,
we present  the mass scales $M_{23}$ and $M_U$, 
and the gauge couplings 
$g_U$  in Table~\ref{B-IIIA12}.
In these models, $M_{23}$ and $M_U$ are respectively 
around $1.2\times 10^{16}$ GeV and $10^{17-18}$ GeV,
and $g_U$ is about $1.4$ in the Type IA1, Type IA2
and Type IIA1 models, and about 2.0 in the Type IIA2 model. 
Because the Type IA1 (IIA1) and Type IA2 (IIA2) models 
respectively have one pair and two pairs of vector-like particles
$XT_i$ and $\overline{XT}_i$, the unified
coupling $g_U$ in the  Type IA1 (IIA1) model is smaller than that
in the  Type IA2 (IIA2) model while $M_U$  in the  Type IA1 (IIA1) model
is larger than that in the  Type IA2 (IIA2) model.

\begin{figure}[htb]
\centering
\includegraphics[width=8cm]{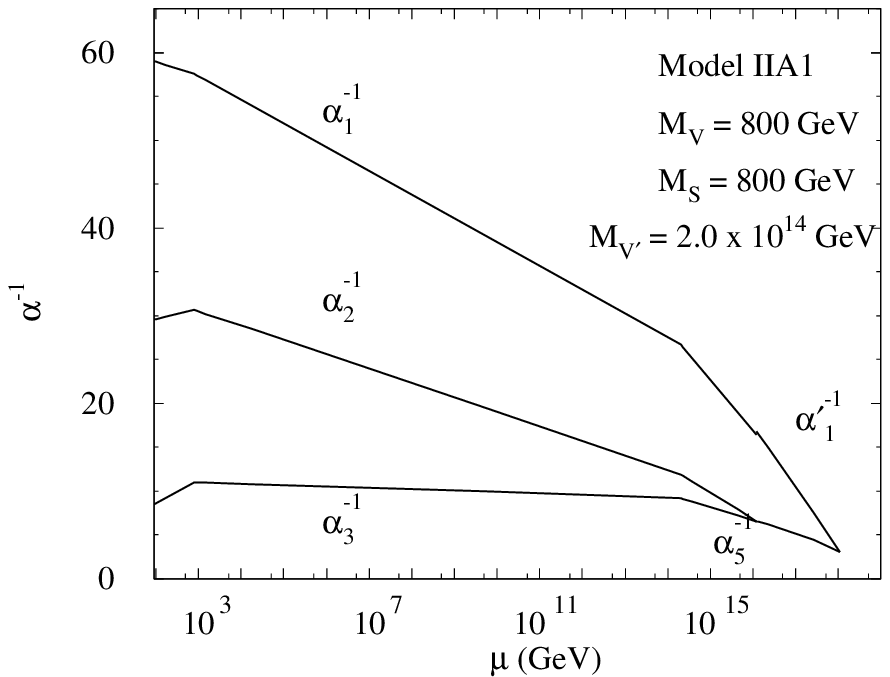}
\includegraphics[width=8cm]{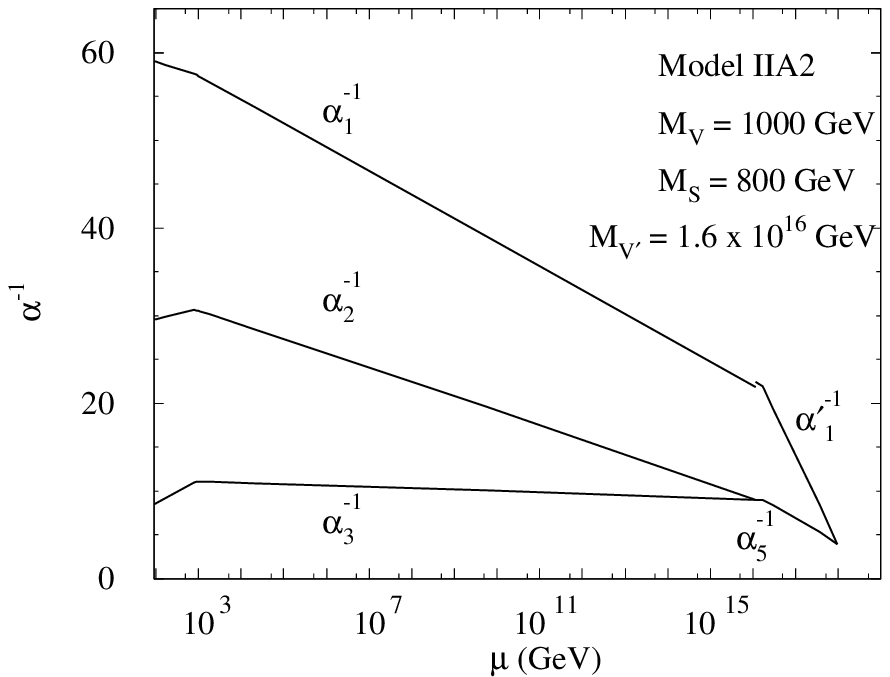}
\caption{The string-scale gauge coupling unification 
in the  Type IIA1 model with $M_V=800~{\rm GeV} $ and 
$M_S=800 ~{\rm GeV} $  (left figure) and in the
Type IIA2  model  with $M_V=1 ~{\rm TeV} $ and 
$M_S=800 ~{\rm GeV} $ (right figure).}
\label{fig:plot98}
\end{figure}

Second, we study the string-scale gauge coupling 
unification at the two-loop level,
and the mass scales $M_{V'}$ are determined from
the condition for string-scale gauge coupling unification
given in Eq.~\ref{String-U}.   In the Type IA1, Type IA2, 
and Type IIA2 models, we choose $M_V=800~{\rm GeV}$
and $M_S=800~{\rm GeV}$. In the Type IIA2 model,
we choose $(M_V, M_S)=(200~{\rm GeV}, ~360~{\rm GeV})$,
$(200~{\rm GeV}, ~1000~{\rm GeV})$, $(1000~{\rm GeV}, ~360~{\rm GeV})$,
and $(1000~{\rm GeV}, ~1000~{\rm GeV})$.
We present  the mass scales $M_{23}$ and $M_{\rm string}$, 
and the gauge couplings 
$g_{\rm string}$  in Table~\ref{C-IIIA12}.
In the Fig.~\ref{fig:plot98}, we present the string-scale 
gauge coupling unification 
in the  Type IIA1 model with $M_V=800~{\rm GeV} $ and 
$M_S=800 ~{\rm GeV} $, and in the 
Type IIA2  model  with $M_V=1 ~{\rm TeV} $ and 
$M_S=800 ~{\rm GeV} $.
We find that $M_{23}$ is about $1.2\times 10^{16}$ GeV in all
the models. In the
Type IA1 and Type IA2 models, 
$M_{\rm string}$ is about $6.35\times 10^{17}~{\rm GeV}$, 
and $g_{\rm string}$ is about $1.2$. In the
Type IIA1 and Type IIA2 models, $M_{\rm string}$ is about 
$1.0\times 10^{18}~{\rm GeV}$, and  $g_{\rm string}$ is
about $2.0$ or larger.
In addition, in the Type IA1, Type IA2 and
Type IIA2 models, the bulk vector-like particles
can be considered as string-scale threshold corrections
since their masses are about $10^{16-17}$ GeV.
While in the Type IIA1 model, the masses for 
the bulk vector-like particles are at the intermediate scale 
$10^{14}$ GeV.


\section{Phenomenological Consequences}

In this Section, we will discuss the phenomenological
consequences in our models.
Similar to the minimal flipped $SU(5)\times U(1)_X$ models,
 the doublet-triplet splitting problem 
can be solved in all of our models in general. Let us comment on
the phenomenological consequences one by one
in the following:

(i) Because GUTs in F-theory are constructed locally,
we may have additional chiral exotic particles or vector-like
particles when we embed such F-theory GUTs into the global
consistent setup. The point is that there may exist additional
seven-branes due to the global consistent conditions, and 
these seven-branes may intersect with the observable 
seven-branes. 

(ii) In the Type IA and Type IIIX2 models where X=A, B, C, D,
we can have $Z1$ set of vector-like particles below the
$1~{\rm TeV}$ scale. Also, in the Type IIA models with
bulk vector-like particles and in the 
 Type IIIX1 models, we have 
 $Z0$ set of vector-like particles below the 
$1~{\rm TeV}$ scale. Thus, these $Z1$ and $Z0$ sets
of vector-like  particles can be produced at the LHC,
and then the corresponding models can be tested.
Moreover, at the low energy,
in the Type IB and Type IIIC3 models, 
we have $Z_2$ set of vector-like particles; 
in the Type IC and Type IIID4 models, we
have $Z_3$ set of vector-like particles; and
in the Type IIB  models with bulk vector-like particles
 and the Type IIID3 model, we have $Z_4$ set of vector-like particles.
The masses for the $Z2$,  $Z3$, or $Z4$ set of 
vector-like particles in these models are around
$10~{\rm TeV}$. Because of the threshold corrections at
the scales $M_{SUSY}$ and $M_{23}$, the masses of these vector-like
particles might be around the $1~{\rm TeV}$ scale,
and then these models could be tested at the LHC as well.
Therefore, all of our models with 
$Z0$, $Z1$,  $Z2$,  $Z3$, or $Z4$ set of
 vector-like particles at the TeV scale might be 
tested at the LHC.
The detail study will be presented elsewhere~\cite{Paper-A}.

(iii) It is well known that 
the lightest CP-even Higgs boson mass in the MSSM
is smaller than about 130 GeV if $M_S$ is smaller than 
$1~{\rm TeV}$, which is a several percents' fine-tuning 
problem in the MSSM.
In all our models with TeV-scale vector-like particles,
we have the vector-like particles $XF$ and $\overline{XF}$.
Then we can introduce the
 following Yukawa interactions between  the MSSM Higgs fields
and these vector-like particles in the 
flipped $SU(5)\times U(1)_X$ models:
\begin{eqnarray}
-{\cal L} &=& y^d_{XF} XF XF h + 
y^u_{XF} \overline{XF} \overline{XF} \overline{h}~,~\,
\end{eqnarray}
where $y^d_{XF}$ and $y^u_{XF}$ are Yukawa couplings. With relatively
large Yukawa couplings $y^d_{XF}$ and $y^u_{XF}$ that are consistent
with the perturbative unification, we can increase the 
 lightest CP-even Higgs boson mass and solve the Higgs mass
problem in the MSSM~\cite{Babu:2008ge, Paper-A}.

(iv) The proton decay via 
dimension-5 operators from Higgsino exchange is suppressed.
Considering proton decay $p  \to e^+ \pi^0$ via dimension-6 
operator from heavy gauge boson exchange,
we obtain the proton life time~\cite{Jiang:2008yf, Ellis:2002vk}  
\begin{eqnarray}
\tau_p \simeq 9.97\times 10^{34} \left({{M_{23}}\over 
{1.18\times 10^{16}{\rm GeV}}}\right)^4
\left( {{1.193}\over {g_{23}}} \right)^4 ~{\rm yr}~,~\,
\end{eqnarray}
where $g_{23}$ is the $SU(3)_C\times SU(2)_L$ unified 
gauge coupling. In all of our models with TeV-scale vector-like
particles (the Type I and Type III models, and Type II
models with bulk vector-like particles), $g_{23}$ is
about 1.2 or larger. In addition, $M_{23}$ can be 
another factor 2/3 smaller due to  threshold 
corrections~\cite{Ellis:2002vk}, thus, our models can definitely 
be tested at the future Hyper-Kamiokande proton decay experiment which
can search the proton life time via $p  \to e^+ \pi^0$ channel 
at least more than $10^{35}$ years~\cite{Nakamura:2003hk}. 
Similar results can be applied to the F-theory $SU(5)$ models
with vector-like particles. Our systematical and comprehensive study
will be presented elsewhere~\cite{Paper-B}.
By the way, the Kaluza-Klein modes of the gauge bosons could  further
enhance the proton decay. However, the details depend on the 
estimations of the bulk Green's functions for the gauge bosons
which have some unknown constants~\cite{Donagi:2008ca}.

(v) From Eq.~(\ref{potgut}), we obtain
that the neutrino masses and mixings can be explained via double seesaw 
mechanism~\cite{Ellis:1992nq}.
Also, the right-handed neutrino Majorana 
masses can be generated via the following dimension-5 operators after we 
integrate out the heavy Kaluza-Klein modes~\cite{Beasley:2008dc, Beasley:2008kw}
\begin{eqnarray}
W = {{y_{ij}^{\prime N}}\over {M_U}} F_i F_j \overline{H} \overline{H}~.~\,
\end{eqnarray}
So the neutrino masses and mixings can be generated via seesaw mechanism as well.
With leptogenesis~\cite{Fukugita:1986hr}, 
we can obtain the observed baryon asymmetry~\cite{Ellis:1992nq}.

(vi) From Eq.~(\ref{spgut}), we can naturally have 
the hybrid inflation where $\Phi$ is the inflaton
 field~\cite{Kyae:2005nv}. The inflation 
scale is related to the scale $M_{23}$. 
Because $M_{23}$ is at least one order smaller
than $M_U$, we solve the monopole problem.
Interestingly, we can 
generate the correct cosmic primordial density 
fluctuations~\cite{Spergel:2006hy}
\begin{eqnarray}
{{\delta \rho}\over {\rho}} \sim \left({{M_{23}}\over 
{g_{23} M_{\rm Pl}}}\right)^2 \sim  1.7\times 10^{-5}
~.~\,
\end{eqnarray}

Therefore, the key question is whether
we can generate $\Phi(\overline{H} H -M_{H^2})$ terms
in Eq.~(\ref{spgut}). Our detail study will be
given elsewhere~\cite{Cos-HLN}. Here let us briefly sketch the
idea. Let us suppose that the $H$ and $\overline{H}$
arise from the intersection of the obsevable seven-branes 
and the seven-brane that wraps a complex
codimension-one surface $S_H$ in $B_3$, and $\Phi$ arises
from the intersection between the seven-brane wrapping
$S_H$ and the seven-brane that wraps a complex
codimension-one surface $S_{\Phi}$ in $B_3$.
Because the curve $\Sigma_H$ is self pinched, we
have the trilinear superpotential $\Phi \overline{H} H $
if the curve on which $\Phi$ is localized passes
the pinched point in $B_3$. In addition,
the term $M_H^2 \Phi $ can be generated via instanton
effects~\cite{Heckman:2008es}. Assume that the volumes for  $S_H$ and 
$S_{\Phi}$ are the same and their compactification
scale is $M_{C2}$, we obtain 
\begin{eqnarray}
M_H ~\simeq~ M_U {\rm exp} ~\left(-{{4\pi^2}\over {g_U^2}} 
{{M_U^4}\over {M_{C2}^4}} \right)~.~\,
\end{eqnarray}
Without fine-tuning, we can choose $M_{C2}=2 M_U$, and
then we can obtain the correct scale for $M_H$.


\section{Discussion and Conclusions}

In this paper, we briefly reviewed the flipped 
$SU(5)\times U(1)$ models and
 the F-theory model building. To separate
the mass scales $M_{23}$ and $M_U$ and realize the
decoupling scenario,  we introduced sets of 
vector-like particles in complete
 $SU(5)\times U(1)$ multiplets at the low energy, 
whose one-loop beta functions satisfy $\Delta b_1 <
  \Delta b_2 = \Delta b_3$. To avoid the Landau pole
problem for the gauge couplings, we can only introduce
five sets of such vector-like particles around the TeV scale.
Moreover, we have systematically
constructed the flipped $SU(5)\times U(1)_X$ models without
 bulk vector-like particles, and the flipped 
$SU(5)\times U(1)_X$ models with bulk vector-like particles.
These vector-like particles can couple to the SM singlet
fields, and obtain suitable masses through Higgs mechanism.
In addition, we considered the gauge coupling 
unification in all of our models without bulk vector-like 
particles, and in the Type IA and Type IIA models with
bulk vector-like particles. We also studied the 
 string-scale gauge coupling unification in the Type III models,
and the Type IA and Type IIA models with bulk vector-like particles.
We showed that the $U(1)_X$ flux contributions to the gauge
couplings preserve the $SU(5)\times U(1)_X$ gauge coupling
unification. We calculated the mass scales $M_{23}$ and $M_U$, 
and the  unified couplings $g_U$.  In the Type IIIX2, Type IIIC3, 
 Type IIID4, Type IA1, Type IA2,
and Type IIA2 models, 
the high-scale or bulk vector-like particles can be 
considered as string-scale threshold corrections 
since their masses are close to the string scale.
We showed that the $Z0$ and $Z1$ sets  of  vector-like particles 
 can have masses below the $1~{\rm TeV}$ scale, and then 
they can be observed at the LHC. Thus, the corresponding
models, which have $Z0$ or $Z1$ sets of vector-like particles 
at about $1~{\rm TeV}$ scale, can be tested at the LHC.

Furthermore, we discussed the phenomenological consequences
of our models. We pointed out that there may exist 
additional chiral exotic particles or vector-like particles 
when we embed our models into the global consistent setup. 
Due to the threshold corrections at the scales $M_S$ and $M_{23}$,
the $Z2$, $Z3$, and $Z4$ sets  of vector-like particles 
 might also have masses below the $1~{\rm TeV}$ scale,
and then the corresponding models with such sets could 
be tested at the LHC as well. In all
our models with TeV-scale vector-like particles, 
the proton decay is within the reach of the future 
Hyper-Kamiokande experiment, the lightest CP-even Higgs boson mass
can be increased due to the Yukawa couplings between the Higgs 
fields and TeV-scale vector-like particles, the neutrino masses 
and mixings can be explained via the double seesaw or seesaw mechanism, 
the observed baryon asymmetry can be obtained through leptogenesis, 
the hybrid inflation can be realized, the monopole problem can
be solved, and
the correct cosmic primodial density fluctuations can be generated.

\begin{acknowledgments}

This research was supported in part 
by the Cambridge-Mitchell Collaboration in Theoretical Cosmology (TL),
by the Natural Science Foundation of China under grant No. 10821504 (TL),
and by the DOE grant DE-FG03-95-Er-40917 (DVN).

\end{acknowledgments}

\appendix

\section{Briefly Review of del Pezzo Surfaces}

The del Pezzo surfaces $dP_n$, where $n=1,~2,~...,~8$, are
defined by blowing up $n$ generic points of 
$\mathbb{P}^{1}\times\mathbb{P}^{1}$ or 
$\mathbb{P}^2$. The homological group
$H_2(dP_n, Z)$ has the generators
\begin{equation}
H,~E_1, ~E_2,~...,~E_n~,~\,
\end{equation}
where $H$ is the hyperplane class for $P^2$, and $E_i$ are the
exceptional divisors at the blowing up points and are
 isomorphic to $\mathbb{P}^{1}$.
 The intersecting numbers of the generators are
\begin{equation}
H\cdot H=1~,\;\;\:E_{i}\cdot E_{j}=-\delta_{ij}~,\;\;\;H\cdot E_{i}=0~.~\,
\end{equation}
The canonical bundle on $dP_{n}$ is given by
\begin{equation}
K_{dP_{n}}=-c_{1}(dP_{n})=-3H+\sum_{i=1}^{n}E_{i}.
\end{equation}
For $n\geq3$,  we can define the generators as follows
\begin{equation}
\alpha_i=E_i-E_{i+1}~,~~~{\rm where}~~i=1,~2,...,~n-1~,~\,
\end{equation}
\begin{equation}
\alpha_n=H-E_1-E_2-E_3~.~\,
\end{equation}
Thus, all the generators $\alpha_i$ is perpendicular to the canonical
class $K_{dP_{n}}$. And
 the intersection products are equal to the negative Cartan
matrix of the Lie algebra $E_n$, and can be considered as simple 
roots. 

The  curves $\Sigma_i$ in $dP_n$ where the particles are localized 
 must be divisors of $S$. And the genus for curve $\Sigma_i$ is
given by 
\begin{equation}
2 g_i-2~=~[\Sigma_i]\cdot ([\Sigma_i]+K_{dP_{k}})~.~\,
\end{equation}

For a line bundle $L$ on the surface $dP_{n}$ with
\begin{equation}
c_{1}(L)=\sum_{i=1}^{n}a_{i}E_{i},
\end{equation}
where $a_{i}a_{j}<0$ for some $i\neq j $,  the K\"ahler form
$J_{dP_{n}}$ can be constructed as follows
\cite{Beasley:2008dc}
\begin{equation}
J_{dP_{k}}=b_0H-\sum_{i=1}^{n}b_{i}E_{i},
\end{equation}
where $\sum_{i=1}^k a_{i}b_{i}=0$ and $b_0 \gg b_{i}>0$. By the
construction, it is easy to see that the line bundle $L$ solves
the BPS equation $J_{dP_k}\wedge c_{1}(L)=0$.

\section{The Vector-Like Particle Curves and 
the Gauge Bundle Assignments in the Models with
One Pair of the Bulk Vector-Like Particles}

In the Type I, Type II, and Type III  models, we can 
introduce one pair of the bulk vector-like particles  
$XT_1$ and $\overline{XT}_1$. Let us choose the 
line bundle $L=\mathcal{O}_{S}(E_{1}-E_{2}+E_4-E_5)^{1/4}$.
Note that $\chi(S, L^4)$ is equal to $-1$, we have 
one pair of the bulk vector-like particles  
$XT_1$ and $\overline{XT}_1$.
We present the curves with homology classes for the
vector-like particles, and the gauge bundle
assignments for each curve in the corresponding
 Type I and Type II models in Table~\ref{Bulk-IA},
and in the corresponding
 Type III models in Table~\ref{Bulk-IB}.

\begin{table}[htb]
\begin{center}
\begin{tabular}{|c|c|c|c|c|c|c|}
\hline
${\rm Model}$ &
${\rm Particles }$ & ${\rm Curve}$ & ${\rm Class}$ & $g_{\Sigma}$ &
$L_{\Sigma}$ & $L_{\Sigma}^{\prime n}$\\\hline
Type I \& II &
$\left(  XF+\overline{XF}\right)  $ & $\Sigma_{XF}\text{ (pinched)}$ &
$3H-E_{4}-E_{5}$ & $1$ & $\mathcal{O}_{\Sigma_{XF}}(p_{45})^{1/4}$ &
$\mathcal{O}_{\Sigma_{XF}}(p_{45})^{-1/4}$\\\hline
Type IA & $\left(  Xl+\overline{Xl}\right)  $ & $\Sigma_{Xl}\text{ (pinched)}$ &
$3H-E_{1}-E_{5}$ & $1$ & $\mathcal{O}_{\Sigma_{Xl}}(p'_{15})^{1/4}$ &
$\mathcal{O}_{\Sigma_{Xl}}(p'_{15})^{-5/4}$\\\hline
Type IB &
$\left(  Xf+\overline{Xf}\right)  $ & $\Sigma_{Xf}\text{ (pinched)}$ &
$3H-E_{1}-E_{5}$ & $1$ & $\mathcal{O}_{\Sigma_{Xf}}(p'_{15})^{1/4}$ &
$\mathcal{O}_{\Sigma_{Xl}}(p'_{15})^{3/4}$\\\hline
Type IC & 
$\left(  Xl+\overline{Xl}\right)  $ & $\Sigma_{Xl}\text{ (pinched)}$ &
$3H-E_{1}-E_{5}$ & $1$ & $\mathcal{O}_{\Sigma_{Xl}}(p'_{15})^{1/4}$ &
$\mathcal{O}_{\Sigma_{Xl}}(p'_{15})^{-5/4}$\\
 & $\left(  Xh+\overline{Xh}\right)  $ & $\Sigma_{Xh}\text{ (pinched)}$ &
$3H-E_{2}-E_{4}$ & $1$ & $\mathcal{O}_{\Sigma_{Xh}}(p'_{42})^{1/4}$ &
$\mathcal{O}_{\Sigma_{Xh}}(p'_{42})^{1/2}$\\\hline
Type IIB & $\left(  Xh+\overline{Xh}\right)  $ & $\Sigma_{Xh}\text{ (pinched)}$ &
$3H-E_{1}-E_{5}$ & $1$ & $\mathcal{O}_{\Sigma_{Xh}}(p'_{15})^{1/4}$ &
$\mathcal{O}_{\Sigma_{Xh}}(p'_{15})^{1/2}$\\\hline
\end{tabular}
\end{center}
\caption{ The vector-like particle curves and the gauge bundle assignments 
for each curve in Type I and Type II models
with one pair of bulk vector-like particles ($XT_1$ and $\overline{XT}_1$).  
Here, $p_{45}=p_4-p_5$, $p'_{15}=p_1'-p_5'$, and
$p'_{42}=p'_4-p'_2$.}
\label{Bulk-IA}
\end{table}

\begin{table}[htb]
\begin{center}
\begin{tabular}{|c|c|c|c|c|c|c|c|}
\hline
Models &${\rm Particles}$ & ${\rm Curve}$ & ${\rm Class}$ & $g_{\Sigma}$ &
$L_{\Sigma}$ & $L_{\Sigma}^{\prime n}$\\\hline
Type III & $\left(  XF+\overline{XF}\right)  $ & 
$\Sigma_{XF}\text{ (pinched)}$ & $ 3H-E_4-E_5 $ & $1$ & 
$\mathcal{O}_{\Sigma_{XF}}(p_{45})^{1/4}$ & $\mathcal{O}_{\Sigma_{XF}}
(p_{45})^{-1/4}$ \\
& $\left(  Xl_i+\overline{Xl}_i\right)  $ & $\Sigma_{Xl_i}\text{ (pinched)}$ &
$3H-E_{1}-E_{2}-E_j$ & $1$ & $\mathcal{O}_{\Sigma_{Xf}}(p^j_{12})^{1/4}$ &
$\mathcal{O}_{\Sigma_{Xf}}(p^j_{12})^{-5/4}$\\\hline
Type IIIA &
$\left(  Xl+\overline{Xl}\right)  $ & $\Sigma_{Xl}\text{ (pinched)}$ &
$3H-E_{1}-E_5$ & $1$ & $\mathcal{O}_{\Sigma_{Xl}}(p'_{15})^{1/4}$ &
$\mathcal{O}_{\Sigma_{Xf}}(p'_{15})^{-5/4}$\\\hline
Type IIIB &
$\left(  Xf+\overline{Xf}\right)  $ & $\Sigma_{Xf}\text{ (pinched)}$ &
$3H-E_{1}-E_5$ & $1$ & $\mathcal{O}_{\Sigma_{Xf}}(p'_{15})^{1/4}$ &
$\mathcal{O}_{\Sigma_{Xf}}(p'_{15})^{3/4}$\\\hline
Type IIID &
$\left(  Xh+\overline{Xh}\right)  $ & $\Sigma_{Xh}\text{ (pinched)}$ &
$3H-E_{1}-E_5$ & $1$ & $\mathcal{O}_{\Sigma_{Xh}}(p'_{15})^{1/4}$ &
$\mathcal{O}_{\Sigma_{Xh}}(p'_{15})^{1/2}$\\\hline
\end{tabular}
\end{center}
\caption{ The vector-like particle curves and the
 gauge bundle assignments for each curve in Type IIIA models 
with one pair of bulk vector-like particles ($XT_1$ and $\overline{XT}_1$).  
Here, $p_{45}=p_4-p_5$, $p'_{15}=p_1'-p_5'$,
$j=i+5$, and $p_{12}^j=p_1^j-p_2^j$
for $j=6,~7,~8$.}
\label{Bulk-IB}
\end{table}

\section{The Vector-Like Particle Curves and the 
Gauge Bundle Assignments in the Models with
Two Pairs of the Bulk Vector-Like Particles}

In the Type I, Type II, and Type III  models, we can also 
introduce two pairs of the bulk vector-like particles  
$XT_i$ and $\overline{XT}_i$. Let us choose the line bundle 
$L=\mathcal{O}_{S}(E_{1}-E_{2}+E_4-E_5+E_6-E_7)^{1/4}$.
Note that $\chi(S, L^4)$ is equal to $-2$, we have 
two pairs of the bulk vector-like particles  
$XT_i$ and $\overline{XT}_i$.
We present the curves with homology classes for the
vector-like particles, and the gauge bundle
assignments for each curve in the corresponding
 Type I and Type II models in Table~\ref{Bulk-IIA},
and in the corresponding
 Type III models in Table~\ref{Bulk-IIB}.

\begin{table}[htb]
\begin{center}
\begin{tabular}{|c|c|c|c|c|c|c|}
\hline
${\rm Model}$ &
${\rm Particles }$ & ${\rm Curve}$ & ${\rm Class}$ & $g_{\Sigma}$ &
$L_{\Sigma}$ & $L_{\Sigma}^{\prime n}$\\\hline
Type I \& II &
$\left(  XF+\overline{XF}\right)  $ & $\Sigma_{XF}\text{ (pinched)}$ &
$3H-E_{4}-E_{5}$ & $1$ & $\mathcal{O}_{\Sigma_{XF}}(p_{45})^{1/4}$ &
$\mathcal{O}_{\Sigma_{XF}}(p_{45})^{-1/4}$\\\hline
Type IA & $\left(  Xl+\overline{Xl}\right)  $ & $\Sigma_{Xl}\text{ (pinched)}$ &
$3H-E_{6}-E_{7}$ & $1$ & $\mathcal{O}_{\Sigma_{Xl}}(p_{67})^{1/4}$ &
$\mathcal{O}_{\Sigma_{Xl}}(p_{67})^{-5/4}$\\\hline
Type IB &
$\left(  Xf+\overline{Xf}\right)  $ & $\Sigma_{Xf}\text{ (pinched)}$ &
$3H-E_{6}-E_{7}$ & $1$ & $\mathcal{O}_{\Sigma_{Xf}}(p_{67})^{1/4}$ &
$\mathcal{O}_{\Sigma_{Xl}}(p_{67})^{3/4}$\\\hline
Type IC & 
$\left(  Xl+\overline{Xl}\right)  $ & $\Sigma_{Xl}\text{ (pinched)}$ &
$3H-E_{6}-E_{7}$ & $1$ & $\mathcal{O}_{\Sigma_{Xl}}(p_{67})^{1/4}$ &
$\mathcal{O}_{\Sigma_{Xl}}(p_{67})^{-5/4}$\\
 & $\left(  Xh+\overline{Xh}\right)  $ & $\Sigma_{Xh}\text{ (pinched)}$ &
$3H-E_{4}-E_{7}$ & $1$ & $\mathcal{O}_{\Sigma_{Xh}}(p'_{47})^{1/4}$ &
$\mathcal{O}_{\Sigma_{Xh}}(p'_{47})^{1/2}$\\\hline
Type IIB & $\left(  Xh+\overline{Xh}\right)  $ & $\Sigma_{Xh}\text{ (pinched)}$ &
$3H-E_{6}-E_{7}$ & $1$ & $\mathcal{O}_{\Sigma_{Xh}}(p_{67})^{1/4}$ &
$\mathcal{O}_{\Sigma_{Xh}}(p_{67})^{1/2}$\\\hline
\end{tabular}
\end{center}
\caption{ The vector-like particle curves and the gauge bundle assignments 
for each curve in Type I and Type II models
with two pairs of bulk vector-like particles ($XT_i$ and $\overline{XT}_i$). 
Here, $p_{45}=p_4-p_5$, $p_{67}=p_6-p_7$, and
$p'_{47} =p'_4-p'_7$.}
\label{Bulk-IIA}
\end{table}


\begin{table}[htb]
\begin{center}
\begin{tabular}{|c|c|c|c|c|c|c|c|}
\hline
Models &${\rm Particles}$ & ${\rm Curve}$ & ${\rm Class}$ & $g_{\Sigma}$ &
$L_{\Sigma}$ & $L_{\Sigma}^{\prime n}$\\\hline
Type III & $\left(  XF+\overline{XF}\right)  $ & 
$\Sigma_{XF}\text{ (pinched)}$ & $ 3H-E_4-E_5 $ & $1$ & 
$\mathcal{O}_{\Sigma_{XF}}(p_{45})^{1/4}$ & $\mathcal{O}_{\Sigma_{XF}}
(p_{45})^{-1/4}$ \\
& $\left(  Xl_1+\overline{Xl}_1\right)  $ & $\Sigma_{Xl_1}\text{ (pinched)}$ &
$3H-E_{1}-E_{5}$ & $1$ & $\mathcal{O}_{\Sigma_{Xf}}(p'_{15})^{1/4}$ &
$\mathcal{O}_{\Sigma_{Xf}}(p'_{15})^{-5/4}$\\
& $\left(  Xl_2+\overline{Xl}_2\right)  $ & $\Sigma_{Xl_2}\text{ (pinched)}$ &
$3H-E_{4}-E_{7}$ & $1$ & $\mathcal{O}_{\Sigma_{Xf}}(p'_{47})^{1/4}$ &
$\mathcal{O}_{\Sigma_{Xf}}(p'_{47})^{-5/4}$\\
& $\left(  Xl_3+\overline{Xl}_3\right)  $ & $\Sigma_{Xl_3}\text{ (pinched)}$ &
$3H-E_{2}-E_{6}$ & $1$ & $\mathcal{O}_{\Sigma_{Xf}}(p'_{62})^{1/4}$ &
$\mathcal{O}_{\Sigma_{Xf}}(p'_{62})^{-5/4}$\\\hline
Type IIIA &
$\left(  Xl+\overline{Xl}\right)  $ & $\Sigma_{Xl}\text{ (pinched)}$ &
$3H-E_{6}-E_7$ & $1$ & $\mathcal{O}_{\Sigma_{Xl}}(p_{67})^{1/4}$ &
$\mathcal{O}_{\Sigma_{Xf}}(p_{67})^{-5/4}$\\\hline
Type IIIB &
$\left(  Xf+\overline{Xf}\right)  $ & $\Sigma_{Xf}\text{ (pinched)}$ &
$3H-E_{6}-E_7$ & $1$ & $\mathcal{O}_{\Sigma_{Xf}}(p_{67})^{1/4}$ &
$\mathcal{O}_{\Sigma_{Xf}}(p_{67})^{3/4}$\\\hline
Type IIID &
$\left(  Xh+\overline{Xh}\right)  $ & $\Sigma_{Xh}\text{ (pinched)}$ &
$3H-E_{6}-E_7$ & $1$ & $\mathcal{O}_{\Sigma_{Xh}}(p_{67})^{1/4}$ &
$\mathcal{O}_{\Sigma_{Xh}}(p_{67})^{1/2}$\\\hline
\end{tabular}
\end{center}
\caption{ The vector-like particle curves and the gauge bundle assignments 
for each curve in Type IIIA models 
with two pairs of bulk vector-like particles ($XT_1$ and $\overline{XT}_1$).  
Here, $p_{45}=p_4-p_5$, $p_{67}=p_6-p_7$,
and $p'_{kl} =p'_k-p'_l$ for $kl=15, ~47,~62$.}
\label{Bulk-IIB}
\end{table}


\section{ Beta Functions for the Bulk Vector-Like Particles 
$XT_i$ and $\overline{XT}_i$}

In the convention of Ref.~\cite{Jiang:2006hf}, we first present 
the one-loop beta functions 
$\Delta b \equiv (\Delta b_1, \Delta b_2, \Delta b_3)$ as
complete supermultiplets from the vector-like particles
$XT_i$ and $\overline{XT}_i$ in the supersymmetric Standard Model
\begin{eqnarray}
\Delta b^{XT + \overline{XT}} =({{39}\over 5}, 3, 3)~.~\,
\end{eqnarray}
Second, we present the  two-loop beta functions from
the vector-like particles $XT_i$ and $\overline{XT}_i$
\begin{eqnarray}
\Delta B^{XT + {{XT^c}}}~=~
\left(
\begin{array}{ccc}
\frac{323}{25}& 15 & \frac{176}{5} \\
 5 & 21 & 16 \\
\frac{22}{5}& 6 & 34 \\
\end{array}\right)~.~\,
\end{eqnarray}

In the flipped $SU(5)\times U(1)_X$ models, we first present 
the one-loop beta functions 
$\Delta b \equiv (\Delta b_1, \Delta b_5)$ as
complete supermultiplets from the vector-like particles
$XT_i$ and $\overline{XT}_i$ 
\begin{eqnarray}
\Delta b^{XT + \overline{XT}} =(8, 3)~.~\,
\end{eqnarray}
Second, we present the two-loop beta functions from
the vector-like particles $XT_i$ and $\overline{XT}_i$
\begin{eqnarray}
\Delta B^{XT + \overline{XT}}~=~
\left(
\begin{array}{cc}
\frac{64}{5}& \frac{576}{5} \\
\frac{24}{5} & \frac{366}{5} \\
\end{array}\right)~.~\,
\end{eqnarray}




\begin{thebibliography}{99}
\itemsep 0.5mm




\bibitem{Buchmuller:2005jr}
  W.~Buchmuller, K.~Hamaguchi, O.~Lebedev and M.~Ratz,
  Phys.\ Rev.\ Lett.\  {\bf 96}, 121602 (2006),
  and references therein.


\bibitem{Lebedev:2006kn}
  O.~Lebedev, H.~P.~Nilles, S.~Raby, S.~Ramos-Sanchez, M.~Ratz, 
P.~K.~S.~Vaudrevange and A.~Wingerter,
  Phys.\ Lett.\  B {\bf 645}, 88 (2007), and references therein.


\bibitem{Kim:2006hw}
  J.~E.~Kim and B.~Kyae,
  Nucl.\ Phys.\  B {\bf 770}, 47 (2007);
  Phys.\ Rev.\  D {\bf 77}, 106008 (2008);
J.~H.~Huh, J.~E.~Kim and B.~Kyae,
  arXiv:0904.1108 [hep-ph].


\bibitem{Braun:2005ux}
  V.~Braun, Y.~H.~He, B.~A.~Ovrut and T.~Pantev,
  Phys.\ Lett.\  B {\bf 618}, 252 (2005);
  JHEP {\bf 0605}, 043 (2006),  and references therein.


\bibitem{Bouchard:2005ag}
  V.~Bouchard and R.~Donagi,
  Phys.\ Lett.\  B {\bf 633}, 783 (2006), 
  and references therein.


\bibitem{Langacker:1991an}
  J.~R.~Ellis, S.~Kelley and D.~V.~Nanopoulos,
  Phys.\ Lett.\ B {\bf 260}, 131 (1991);
  P.~Langacker and M.~X.~Luo,
  Phys.\ Rev.\ D {\bf 44}, 817 (1991);
  U.~Amaldi, W.~de Boer and H.~Furstenau,
  Phys.\ Lett.\ B {\bf 260}, 447 (1991).


\bibitem{Dienes:1996du}
For a review, see K.~R.~Dienes,
  Phys.\ Rept.\  {\bf 287}, 447 (1997), and references therein.


\bibitem{Horava:1995qa}
  P.~Horava and E.~Witten,
  Nucl.\ Phys.\ B {\bf 460}, 506 (1996).


\bibitem{Witten:1996mz}
  E.~Witten,
  Nucl.\ Phys.\ B {\bf 471}, 135 (1996).


\bibitem{AEHN}
I.~Antoniadis, J.~R.~Ellis, J.~S.~Hagelin and D.~V.~Nanopoulos,
Phys.\ Lett.\ B {\bf 205} (1988) 459;
Phys.\ Lett.\ B {\bf 208} (1988) 209 [Addendum-ibid.\ B {\bf 213}
(1988) 562];
Phys.\ Lett.\ B {\bf 231} (1989) 65.


\bibitem{Faraggi:1989ka}
  A.~E.~Faraggi, D.~V.~Nanopoulos and K.~J.~Yuan,
  Nucl.\ Phys.\ B {\bf 335}, 347 (1990).


\bibitem{Antoniadis:1990hb}
  I.~Antoniadis, G.~K.~Leontaris and J.~Rizos,
  Phys.\ Lett.\ B {\bf 245}, 161 (1990).


\bibitem{LNY}
  J.~L.~Lopez, D.~V.~Nanopoulos and K.~J.~Yuan,
  Nucl.\ Phys.\ B {\bf 399}, 654 (1993);
  D.~V.~Nanopoulos,
  hep-ph/0211128.


\bibitem{Cleaver:2001ab}
  G.~B.~Cleaver, A.~E.~Faraggi, D.~V.~Nanopoulos and J.~W.~Walker,
  Nucl.\ Phys.\  B {\bf 620}, 259 (2002), and references therein.



\bibitem{Berkooz:1996km}
  M.~Berkooz, M.~R.~Douglas and R.~G.~Leigh,
  Nucl.\ Phys.\  B {\bf 480}, 265 (1996).

\bibitem{Ibanez:2001nd}
  L.~E.~Ibanez, F.~Marchesano and R.~Rabadan,
  JHEP {\bf 0111}, 002 (2001).

\bibitem{Blumenhagen:2001te}
  R.~Blumenhagen, B.~Kors, D.~Lust and T.~Ott,
  Nucl.\ Phys.\  B {\bf 616}, 3 (2001).

\bibitem{CSU}
M.~Cveti\v c, G.~Shiu and A.~M.~Uranga, Phys.\ Rev.\ Lett.\  {\bf
87}, 201801 (2001);
M.~Cveti\v c, G.~Shiu and A.~M.~Uranga, Nucl.\ Phys.\ B {\bf 615},
3 (2001).


\bibitem{Cvetic:2002pj}
  M.~Cveti\v c, I.~Papadimitriou and G.~Shiu,
  Nucl.\ Phys.\ B {\bf 659}, 193 (2003)
  [Erratum-ibid.\ B {\bf 696}, 298 (2004)].


\bibitem{CLL}
M.~Cveti\v c, T.~Li and T.~Liu,
Nucl.\ Phys.\ B {\bf 698}, 163 (2004).
  M.~Cveti\v c, P.~Langacker, T.~Li and T.~Liu,
  Nucl.\ Phys.\ B {\bf 709}, 241 (2005).


\bibitem{Chen:2005ab}
  C.-M.~Chen, G.~V.~Kraniotis, V.~E.~Mayes, D.~V.~Nanopoulos and J.~W.~Walker,
  Phys.\ Lett.\ B {\bf 611}, 156 (2005);
  Phys.\ Lett.\  B {\bf 625}, 96 (2005).


\bibitem{Chen:2005mj}
  C.~M.~Chen, T.~Li and D.~V.~Nanopoulos,
  Nucl.\ Phys.\ B {\bf 732}, 224 (2006).


\bibitem{Blumenhagen:2005mu}
  R.~Blumenhagen, M.~Cvetic, P.~Langacker and G.~Shiu,
  Ann.\ Rev.\ Nucl.\ Part.\ Sci.\  {\bf 55}, 71 (2005),
and references therein.


\bibitem{Dijkstra:2004ym}
  T.~P.~T.~Dijkstra, L.~R.~Huiszoon and A.~N.~Schellekens,
  Phys.\ Lett.\  B {\bf 609}, 408 (2005).


\bibitem{Dijkstra:2004cc}
  T.~P.~T.~Dijkstra, L.~R.~Huiszoon and A.~N.~Schellekens,
  Nucl.\ Phys.\  B {\bf 710}, 3 (2005), and references therein.

\bibitem{Blumenhagen:2008zz}
  R.~Blumenhagen, V.~Braun, T.~W.~Grimm and T.~Weigand,
  Nucl.\ Phys.\  B {\bf 815}, 1 (2009).


\bibitem{Blumenhagen:2006xt}
  R.~Blumenhagen, M.~Cvetic and T.~Weigand,
  Nucl.\ Phys.\  B {\bf 771}, 113 (2007);
L.~E.~Ibanez and A.~M.~Uranga,
  JHEP {\bf 0703}, 052 (2007);
R.~Blumenhagen, M.~Cvetic, D.~Lust, R.~Richter and T.~Weigand,
  Phys.\ Rev.\ Lett.\  {\bf 100}, 061602 (2008).


\bibitem{Chen:2007px}
  C.~M.~Chen, T.~Li, V.~E.~Mayes and D.~V.~Nanopoulos,
  Phys.\ Lett.\  B {\bf 665}, 267 (2008).


\bibitem{Chen:2007zu}
  C.~M.~Chen, T.~Li, V.~E.~Mayes and D.~V.~Nanopoulos,
  Phys.\ Rev.\  D {\bf 77}, 125023 (2008).

\bibitem{Gogoladze:2005az}
  I.~Gogoladze, T.~Li, Y.~Mimura and S.~Nandi,
  Phys.\ Lett.\  B {\bf 622}, 320 (2005);
  Phys.\ Rev.\  D {\bf 72}, 055006 (2005).




\bibitem{Vafa:1996xn}
  C.~Vafa,
  Nucl.\ Phys.\  B {\bf 469}, 403 (1996).


\bibitem{Donagi:2008ca}
  R.~Donagi and M.~Wijnholt,
  arXiv:0802.2969 [hep-th].


\bibitem{Beasley:2008dc}
  C.~Beasley, J.~J.~Heckman and C.~Vafa,
  JHEP {\bf 0901}, 058 (2009).


\bibitem{Beasley:2008kw}
  C.~Beasley, J.~J.~Heckman and C.~Vafa,
  JHEP {\bf 0901}, 059 (2009).



\bibitem{Donagi:2008kj}
  R.~Donagi and M.~Wijnholt,
  arXiv:0808.2223 [hep-th].


\bibitem{Heckman:2008es}
  J.~J.~Heckman, J.~Marsano, N.~Saulina, S.~Schafer-Nameki and C.~Vafa,
  arXiv:0808.1286 [hep-th].


\bibitem{Marsano:2008jq}
  J.~Marsano, N.~Saulina and S.~Schafer-Nameki,
  arXiv:0808.1571 [hep-th].


\bibitem{Marsano:2008py}
  J.~Marsano, N.~Saulina and S.~Schafer-Nameki,
  arXiv:0808.2450 [hep-th].


\bibitem{Heckman:2008qt}
  J.~J.~Heckman and C.~Vafa,
  arXiv:0809.1098 [hep-th].


\bibitem{Font:2008id}
  A.~Font and L.~E.~Ibanez,
  JHEP {\bf 0902}, 016 (2009).


\bibitem{Braun:2008pz}
  A.~P.~Braun, A.~Hebecker, C.~Ludeling and R.~Valandro,
  Nucl.\ Phys.\  B {\bf 815}, 256 (2009).


\bibitem{Heckman:2008qa}
  J.~J.~Heckman and C.~Vafa,
  arXiv:0811.2417 [hep-th].


\bibitem{Jiang:2008yf}
  J.~Jiang, T.~Li, D.~V.~Nanopoulos and D.~Xie,
  arXiv:0811.2807 [hep-th].

\bibitem{Collinucci:2008zs}
  A.~Collinucci,
  arXiv:0812.0175 [hep-th].


\bibitem{Blumenhagen:2008aw}
  R.~Blumenhagen,
  Phys.\ Rev.\ Lett.\  {\bf 102}, 071601 (2009).


\bibitem{Heckman:2008jy}
  J.~J.~Heckman, A.~Tavanfar and C.~Vafa,
  arXiv:0812.3155 [hep-th].

\bibitem{Bourjaily:2009vf}
  J.~L.~Bourjaily,
  arXiv:0901.3785 [hep-th].



\bibitem{Hayashi:2009ge}
  H.~Hayashi, T.~Kawano, R.~Tatar and T.~Watari,
  arXiv:0901.4941 [hep-th].


\bibitem{Andreas:2009uf}
  B.~Andreas and G.~Curio,
  arXiv:0902.4143 [hep-th].


\bibitem{Chen:2009me}
  C.~M.~Chen and Y.~C.~Chung,
  arXiv:0903.3009 [hep-th].


\bibitem{Heckman:2009bi}
  J.~J.~Heckman, G.~L.~Kane, J.~Shao and C.~Vafa,
  arXiv:0903.3609 [hep-ph].

\bibitem{Donagi:2009ra}
  R.~Donagi and M.~Wijnholt,
  arXiv:0904.1218 [hep-th].

\bibitem{Bouchard:2009bu}
  V.~Bouchard, J.~J.~Heckman, J.~Seo and C.~Vafa,
  arXiv:0904.1419 [hep-ph].


\bibitem{Randall:2009dw}
  L.~Randall and D.~Simmons-Duffin,
  arXiv:0904.1584 [hep-ph].


\bibitem{Heckman:2009de}
  J.~J.~Heckman and C.~Vafa,
  arXiv:0904.3101 [hep-th].


\bibitem{Marsano:2009ym}
  J.~Marsano, N.~Saulina and S.~Schafer-Nameki,
  arXiv:0904.3932 [hep-th].

\bibitem{Bourjaily:2009ci}
  J.~L.~Bourjaily,
  arXiv:0905.0142 [hep-th].





\bibitem{smbarr} S. M. Barr,
Phys.\ Lett.\ B {\bf 112}, 219 (1982).


\bibitem{dimitri}
J.~P.~Derendinger, J.~E.~Kim and D.~V.~Nanopoulos,
Phys.\ Lett.\ B {\bf 139}, 170 (1984).

\bibitem{AEHN-0}
I.~Antoniadis, J.~R.~Ellis, J.~S.~Hagelin and D.~V.~Nanopoulos,
Phys.\ Lett.\ B {\bf 194}, 231 (1987).


\bibitem{Lopez:1995cs}
  J.~L.~Lopez and D.~V.~Nanopoulos,
  Phys.\ Rev.\ Lett.\  {\bf 76}, 1566 (1996).


\bibitem{Jiang:2006hf}
  J.~Jiang, T.~Li and D.~V.~Nanopoulos,
  Nucl.\ Phys.\  B {\bf 772}, 49 (2007).


\bibitem{Nakamura:2003hk}
  K.~Nakamura,
  Int.\ J.\ Mod.\ Phys.\  A {\bf 18}, 4053 (2003).


\bibitem{Conlon:2009xf}
  J.~P.~Conlon,
  JHEP {\bf 0904}, 059 (2009).


\bibitem{Huang:2004ui}
  C.~S.~Huang, J.~Jiang and T.~Li,
  Nucl.\ Phys.\  B {\bf 702}, 109 (2004), and the references therein.


\bibitem{Amsler:2008zz}
  C.~Amsler {\it et al.}  [Particle Data Group],
  Phys.\ Lett.\  B {\bf 667}, 1 (2008).

\bibitem{Paper-A}
 Y.~J.~Huo, J.~Jiang, T.~Li, D.~V.~Nanopoulos, C.~L.~Tong, 
in preparation.


\bibitem{Babu:2008ge}
  K.~S.~Babu, I.~Gogoladze, M.~U.~Rehman and Q.~Shafi,
  Phys.\ Rev.\  D {\bf 78}, 055017 (2008).



\bibitem{Ellis:2002vk}
  J.~R.~Ellis, D.~V.~Nanopoulos and J.~Walker,
  Phys.\ Lett.\  B {\bf 550}, 99 (2002).


\bibitem{Paper-B}
  T.~Li, D.~V.~Nanopoulos and J.~Walker, in preparation.


\bibitem{Ellis:1992nq}
  J.~R.~Ellis, D.~V.~Nanopoulos and K.~A.~Olive,
  Phys.\ Lett.\  B {\bf 300}, 121 (1993);
J.~R.~Ellis, J.~L.~Lopez, D.~V.~Nanopoulos and K.~A.~Olive,
  Phys.\ Lett.\  B {\bf 308}, 70 (1993).


\bibitem{Fukugita:1986hr}
  M.~Fukugita and T.~Yanagida,
  Phys.\ Lett.\  B {\bf 174}, 45 (1986).


\bibitem{Kyae:2005nv}
  B.~Kyae and Q.~Shafi,
  Phys.\ Lett.\  B {\bf 635}, 247 (2006).


\bibitem{Spergel:2006hy}
  D.~N.~Spergel {\it et al.}  [WMAP Collaboration],
  Astrophys.\ J.\ Suppl.\  {\bf 170}, 377 (2007).


\bibitem{Cos-HLN}
S.~Hu, T.~Li and D.~V.~Nanopoulos, in preparation.






\end{thebibliography}
\end{document}